\documentclass[sn-standardnature]{sn-jnl}

\newcommand{\lat}{\textit{Fermi}/LAT}
\newcommand{\gbm}{\textit{Fermi}/GBM}
\newcommand{\xrt}{\textit{Swift}/XRT}
\newcommand{\bat}{\textit{Swift}/BAT}

\usepackage{comment}
\usepackage{amsmath}
\usepackage{graphicx}
\usepackage{subcaption}
\usepackage{caption}

\usepackage{multibib}
\newcites{supp}{References}

\jyear{2022}%
\begin{document}

\title{GeV emission from a compact binary merger}

\author*[1,2]{\fnm{Alessio} \sur{Mei}}\email{alessio.mei@gssi.it}

\author[1,2]{\fnm{Biswajit} \sur{Banerjee}}

\author[1,2]{\fnm{Gor} \sur{Oganesyan}}

\author[3,6]{\fnm{Om Sharan} \sur{Salafia}}

\author[4,5]{\fnm{Stefano} \sur{Giarratana}}

\author[1,2]{\fnm{Marica} \sur{Branchesi}}
\author[6]{\fnm{Paolo} \sur{D'Avanzo}}
\author[6]{\fnm{Sergio} \sur{Campana}}
\author[3,6]{\fnm{Giancarlo} \sur{Ghirlanda}}
\author[1,2]{\fnm{Samuele} \sur{Ronchini}}
\author[7]{\fnm{Amit} \sur{Shukla}}
\author[7]{\fnm{Pawan} \sur{Tiwari}}

\affil[1]{ \orgname{Gran Sasso Science Institute}, \orgaddress{\street{Viale F. Crispi 7}, \city{L'Aquila (AQ)}, \postcode{I-67100}, \country{Italy}}}

\affil[2]{ \orgname{INFN - Laboratori Nazionali del Gran Sasso}, \orgaddress{ \city{L'Aquila (AQ)}, \postcode{I-67100}, \country{Italy}}}

\affil[3]{\orgname{Universit\`{a} degli Studi di Milano-Bicocca}, \orgaddress{piazza dell'Ateneo Nuovo 1, \city{Milano (MI)}, \postcode{I-20126}, \country{Italy}}} 

\affil[3]{\orgname{INFN – Sezione di Milano-Bicocca}, \orgaddress{Piazza della Scienza 2, \city{Milano (MI)}, \postcode{I-20126}, \country{Italy}}}

\affil[4]{ \orgname{Department of Physics and Astronomy, University of Bologna}, \orgaddress{ \city{via Gobetti 93/2, Bologna}, \postcode{40129}, \country{Italy}}}

\affil[5]{ \orgname{INAF - Istituto di Radioastronomia}, \orgaddress{ \city{via Gobetti 101, Bologna}, \postcode{40129}, \country{Italy}}}

\affil[6]{ \orgname{INAF - Osservatorio Astronomico di Brera}, \orgaddress{ \city{via E.\ Bianchi 46, Merate (LC)}, \postcode{I-23807}, \country{Italy}}}

\affil[7]{ \orgname{Department of Astronomy, Astrophysics and Space Engineering, Indian Institute of Technology Indore},\orgaddress{ \city{Khandwa Road, Simrol, Indore}, \postcode{453552}, \country{India}}}

\abstract{
An energetic $\rm \gamma$-ray burst (GRB), GRB 211211A, was observed on December 11, 2021 \cite{2021GCN,gbm2021GCN}. Despite its long duration, typically associated with bursts produced by the collapse of massive stars, the observation of an optical-infrared kilonova 
points to a compact binary merger origin \cite{Rastinejad2022}. 
Here we report observations of a significant ($\rm >5 \sigma$) transient-like emission in the high energy $\rm \gamma$-rays (HE; $\rm E>0.1$\,GeV) starting $10^3$\,s after the burst. After an initial phase with a roughly constant flux ($\rm \sim 5\times 10^{-10}\ erg\ s^{-1}\ cm^{-2}$) lasting $\sim 2\times 10^4$\,s, the flux started decreasing and soon went undetected. 
Our detailed modelling of public and dedicated multi-wavelength observations demonstrates that GeV emission from GRB 211211A is in excess with respect to the flux predicted by the state-of-the-art afterglow model at such late time.
We explore the possibility that the GeV excess is inverse Compton emission due to the interaction of a late-time, low-power jet with an external source of photons, and find that kilonova emission can provide the seed photons. Our results open new perspectives for observing binary neutron star mergers.}

\maketitle


$\rm \gamma$-ray bursts (GRBs) are extra-galactic transients which release an enormous amount of (isotropic equivalent) energy, of the order of $\rm 10^{52}-10^{54}$ erg. The initial highly variable $\rm \gamma-$ray radiation (called `prompt emission') is a short-lived burst ($\rm 0.1-10^{3}$ s) in the $\rm \gamma$-ray band (10 keV - 10 MeV) originating from internal dissipation of energy within an ultra-relativistic jet \citep{narayan1992, rees1994,sari1996}. The propagation of the GRB jet into the circum-burst medium produces a shock wave which gives rise to a multi-wavelength, longer-lived afterglow emission due to synchrotron radiation from non-thermal electrons \citep{Paczynski1993,Sari1999}. 
GRBs are classified into two classes based on the duration of their prompt emission: `long' (lGRBs, longer than 2 sec) and `short' (sGRBs, shorter than 2 sec) \cite{Kouveliotou1993}. Over the last decades, extensive GRB studies demonstrated that their spectral properties, afterglow emission and host galaxies are consistent with two types of progenitors \cite{berger2014}. While a number of supernovae (SNe) of the Ib/c class detected in association with long GRBs \cite{woosley2006} revealed massive star collapse progenitors, the short GRB 170817A detected in close temporal association with the gravitational wave (GW) signal GW170817 \cite{abbott2017A,abbott2017B,Abbott2017C} showed that binary neutron star (BNS) mergers are the progenitors of short GRBs.
A key feature of compact object mergers involving at least one neutron star is the kilonova, an optical-infrared transient powered by radioactive decay of unstable heavy isotopes synthetized by rapid neutron capture by nuclei within the expanding neutron-rich merger ejecta \cite{Li1998,Metzger2010}. Signatures of the presence of such an emission have been observed following several sGRBs \cite{Tanvir2013,Berger2013,Troja2019,Ascenzi2019,Jin2020,rossi2020}, and the first spectroscopic confirmation and detailed multi-wavelength characterization came with the AT2017gfo kilonova after GW170817 \cite{abbott2017B,Pian2017,Smartt2017,Drout2017}.
\begin{figure*}
	\centering 
	\includegraphics[width=1\textwidth]{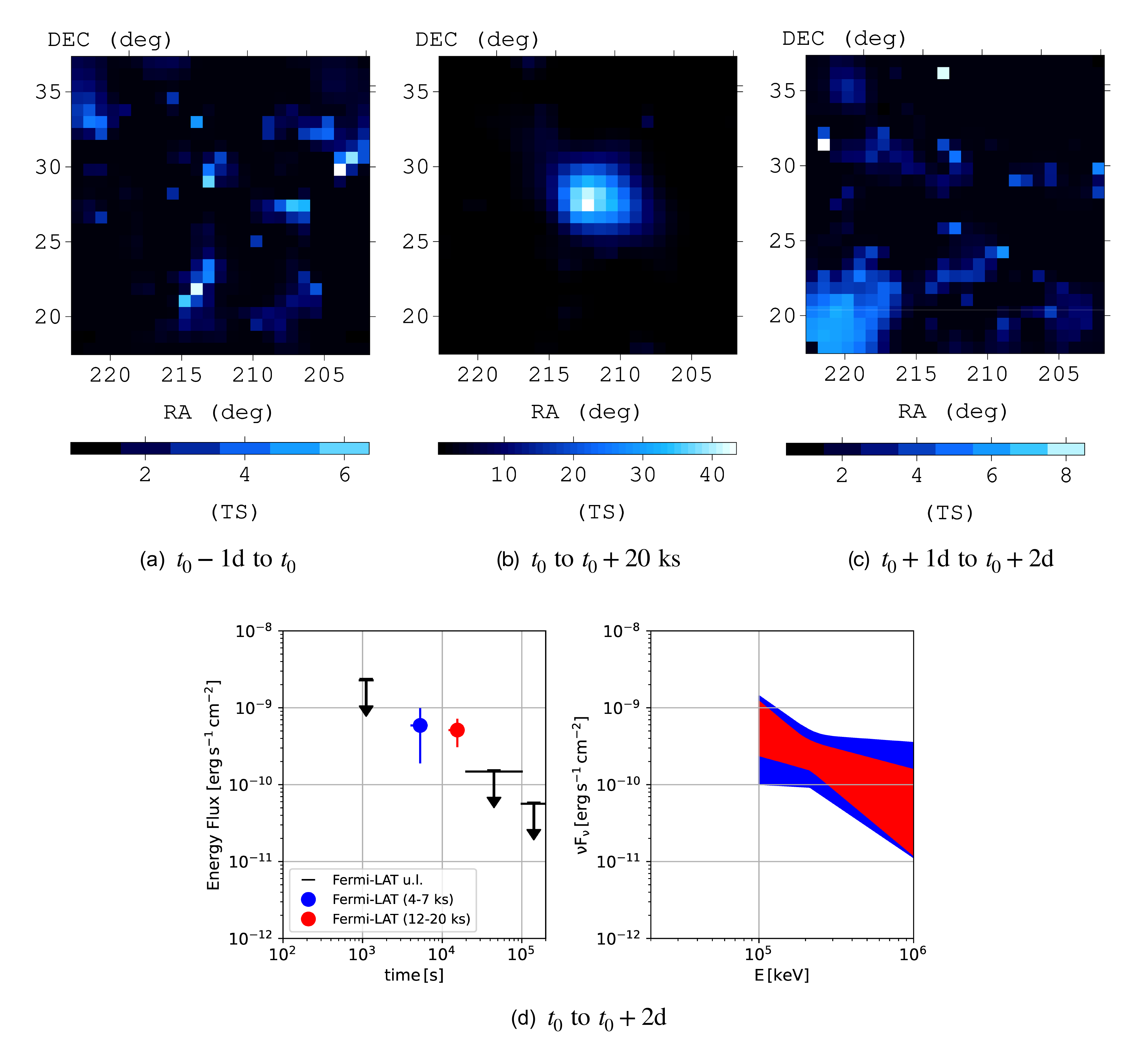} 
	\caption{}
	\label{fig:TSmaps}
\end{figure*}\\
On December 11, 2021 at 13:09:59 Universal Time (UT) a burst triggered the Burst Alert Telescope (BAT) on-board the \textit{Swift} satellite \cite{2021GCN} and the Gamma-ray Burst Monitor (GBM) on-board the \textit{Fermi} satellite \cite{gbm2021GCN}. The GRB 211211A showed duration and properties typical of long GRBs, but right after its detection, several outstanding features came into sight. Deep optical observations revealed that this source is strongly offset with respect to the center of its host galaxy, placed at redshift $z=0.076$ (350 Mpc) \cite{Rastinejad2022}. A kilonova emission was discovered in the optical/NIR band in temporal and spatial coincidence with the burst \cite{Rastinejad2022}. In addition, the $\gamma$-ray precursor anticipating the prompt emission shows signatures of Quasi-Periodic Oscillations (QPOs, \cite{Xiao2022}). Despite its long duration as estimated by \textit{Swift} and \textit{Fermi}, these interesting discoveries are pointing toward a compact binary merger progenitor, 
strongly challenging the above short/long GRB progenitor dichotomy and opening a new door to a more complex classification scheme for GRBs.\\
Here, we report another discovery for the GRB 211211A; the detection, in the \textit{Fermi} Large Area Telescope (LAT) data, of a significant ($>5\sigma$) emission with photon energies between 0.1-1\,GeV (Fig.~\ref{fig:TSmaps}). 
The emission showed up significantly ($\geq3 \sigma$) about 6\,ks after the burst and remained constant for about other 14\,ks.
It is characterised by a relatively soft spectrum with power law photon index of $\rm 2.9 \pm 0.4$. We measure a flux of $\rm  (5.21\pm 1.52)\times 10^{-10}\ erg\ s^{-1}\ cm^{-2}$ between 0.1 and 1 GeV (integrated between 1 and 20\,ks after the trigger time) which, given the source distance, corresponds to a luminosity of $\rm (7.4\pm2.2)\times 10^{45}\ erg\ s^{-1}$, particularly low compared to GRBs observed so far at similar times and frequencies by \lat\ \cite{ajello2019}, as shown in Fig.~\ref{fig:LAT_catalog}. This intrinsically faint emission would be hardly detected at distances larger than 350 Mpc. Such a late time emission in {\it Fermi}/LAT data is not present in any other GRBs closer than 350 Mpc (see Methods), and it has never been reported for short GRBs at any distances (Fig.~\ref{fig:LAT_catalog}). For GW170817, no {\it Fermi}/LAT detection was reported on timescales of minutes, hours, and days after the BNS merger \cite{2018ApJ...861...85A}, making the GeV emission of GRB 211211A the first ever high energy component observed in association with a compact binary merger event (see Methods). Recently, GeV emission has been detected from a binary system containing one massive star and one compact object \citep{xingxing2020}.
\begin{figure*}
	\centering 
	\includegraphics[width=1\textwidth]{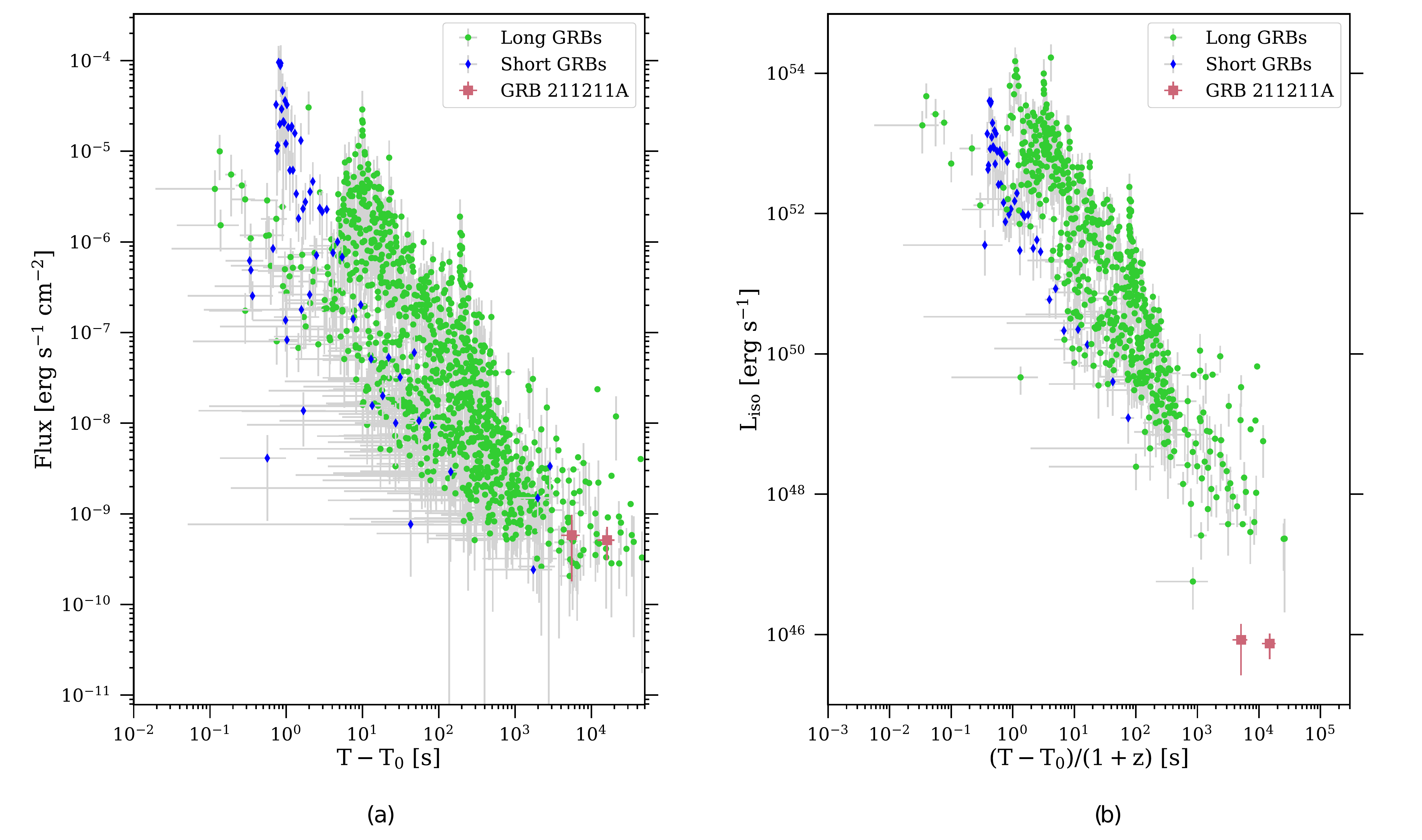} 
	\caption{}
  \label{fig:LAT_catalog}
\end{figure*}\\
We have followed up GRB 211211A with the High Throughput X-ray Spectroscopy Mission (\textit{XMM}-Newton) in soft X-rays (0.3 - 10 keV) 9.6 and 51 days after GRB trigger time and we have obtained deep upper limits of $\rm 10^{-14} \, erg \, s^{-1} \, cm^{-2}$ and $\rm 5\times 10^{-15} \, erg \, s^{-1} \, cm^{-2}$, respectively. A search for late radio afterglow emission was performed 35, 39 and 77 days post-burst with the Karl G. Jansky Very Large Array (VLA) at 3, 6 and 10 GHz frequencies. We did not detect any emission also at these frequencies. To fully characterize the afterglow emission of this source, we enriched our dataset with publicly available data from \xrt, \textit{Swift}/UVOT, and photometry from ground-based optical/IR telescopes (see Methods for details). We model radio-to-GeV observations within the standard afterglow scenario \cite{sari1996}, including synchrotron and synchrotron self-Compton (SSC) radiation from shock-accelerated circum-burst medium electrons. We also include a simple one-component model for the kilonova emission (see Methods). The model fit is in good agreement with the optical and X-ray light curves, including the well constrained spectral shape of the soft X-ray emission and the very late epoch upper limits obtained with the VLA and \textit{XMM}-Newton (Fig. \ref{fig:grb1_sed_lc}).
\begin{figure*}
\centering 
\includegraphics[width=\textwidth]{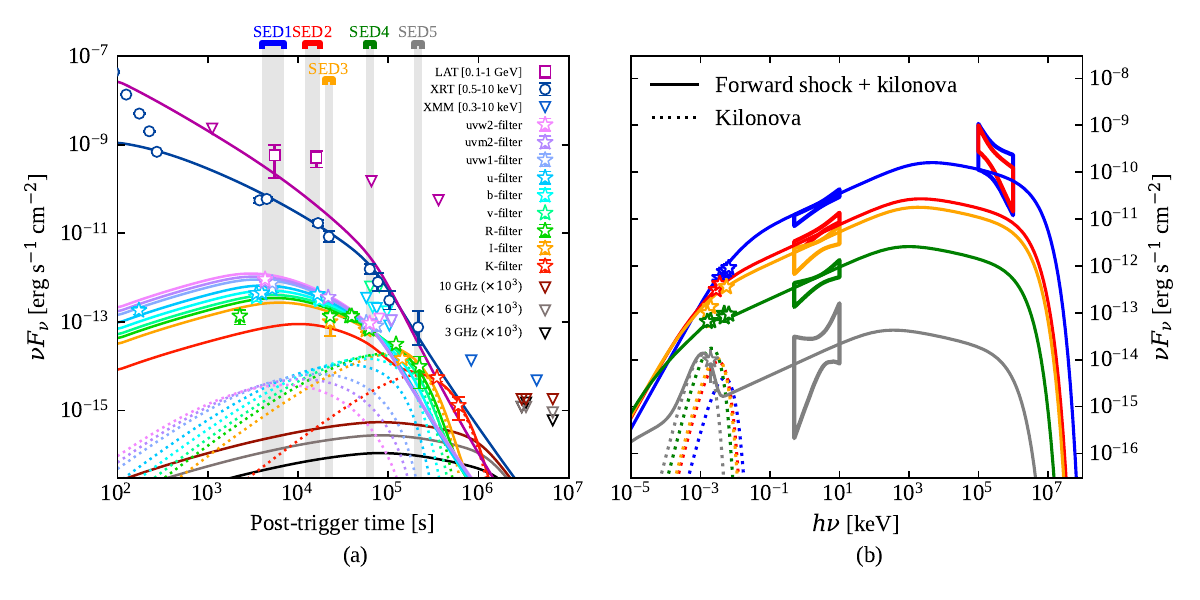} 
\caption{}
\label{fig:grb1_sed_lc}
\end{figure*}\\
\begin{figure*}
\centering 
\includegraphics[width=\textwidth]{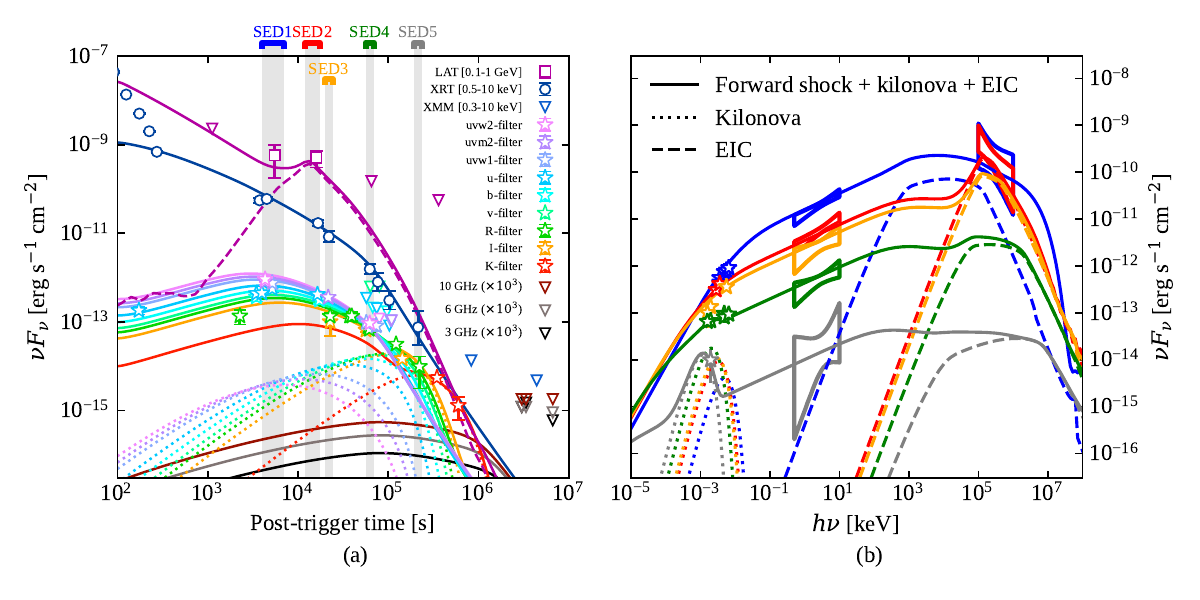} 
\caption{}
\label{fig:grb1_sed_lc_EIC}
\end{figure*}\\
The best fit parameters of the afterglow suggest that the GRB jet is highly collimated (aperture angle $\theta_\mathrm{j} \sim 1.0_{-0.3}^{+0.5}\,\rm deg$, 90\% credible level) and it propagates in a rarefied circum-burst medium with a homogeneous number density $n \le 8\times 10^{-5} \, \rm  cm^{-3}$ (95\% credible upper limit), in agreement with what would be expected given the offset between this GRB and its host galaxy center. The jet's total kinetic energy (corrected for collimation) $E_{\rm jet} = 1.0^{+6.0}_{-0.9}\times 10^{50}\, \rm erg$ is consistent with the amount of energy disposed in the jet formed from the BNS merger of GW170817 (e.g. \cite{salafia2021}).\\
Despite the good spectral and timing description of the optical and X-ray data by the combined standard afterglow and the kilonova models, the high-energy emission component at late times remains in excess with respect to the standard forward shock radiation. While the first epoch of the \lat\ observation (4-7 ks) is consistent with the synchrotron component of the forward shock, the second one (between 12-20 ks) is in clear excess over the expected power law decay of the afterglow ($\propto t^{-1}$, see Fig.~\ref{fig:grb1_sed_lc}, panel a). Such a late excess could originate from the External Inverse Compton (EIC) process, that is, from photons produced externally with respect to the GRB jet that are up-scattered by electrons within the latter. A possible source of external photons is the kilonova emission. Its thermal emission spectrum, Comptonized by the jet electrons,  would account for the relatively soft observed GeV spectrum and would not contaminate the soft X-ray which is well described by the synchrotron emission. Assuming the kilonova photons to be the seed photons for the EIC, we conclude that the relativistic electrons from the forward shock  responsible for the multi-wavelength emission cannot reproduce the observed luminosity of the GeV component.This is because at $\sim10^4$ s the dominant  photons would come from the forward shock, but the syncrothron self Compton peaks at higher energies and its contribution is negligible to the GeV emission. On the other hand, a source of hot electrons closer to the kilonova ejecta can account for the EIC component. We invoke the presence of a low-power jet ejected at late times, whose electrons are capable of producing the amount of GeV emission without over-shining the overall multi-wavelength afterglow emission by the synchrotron radiation.\\
If the hot electrons are placed above the kilonova photosphere (de-beamed scenario, Fig.~\ref{fig:draw}, panel a) the EIC process is inefficient and it requires a very low magnetization of the jet to not outshine the observed afterglow emission by the Synchrotron radiation. Instead, we show that a scenario where such electrons reside below the kilonova photosphere at $t=10^4$ s (beamed scenario, Fig.~\ref{fig:draw}, panel b) leads to beamed seed photons in the comoving frame of the jet, hence explaining the observed GeV emission by inverse Compton scattering of the kilonova photons without extreme requirements on the magnetic field strength (see Methods for details). GeV emission 
from up-scattered kilonova photons 
opens new perspectives in detecting kilonova emission at high energies and it represents another relevant electromagnetic counterpart of gravitational-waves from compact binary mergers.
\begin{figure*}
	\centering 
	\includegraphics[width=1\textwidth]{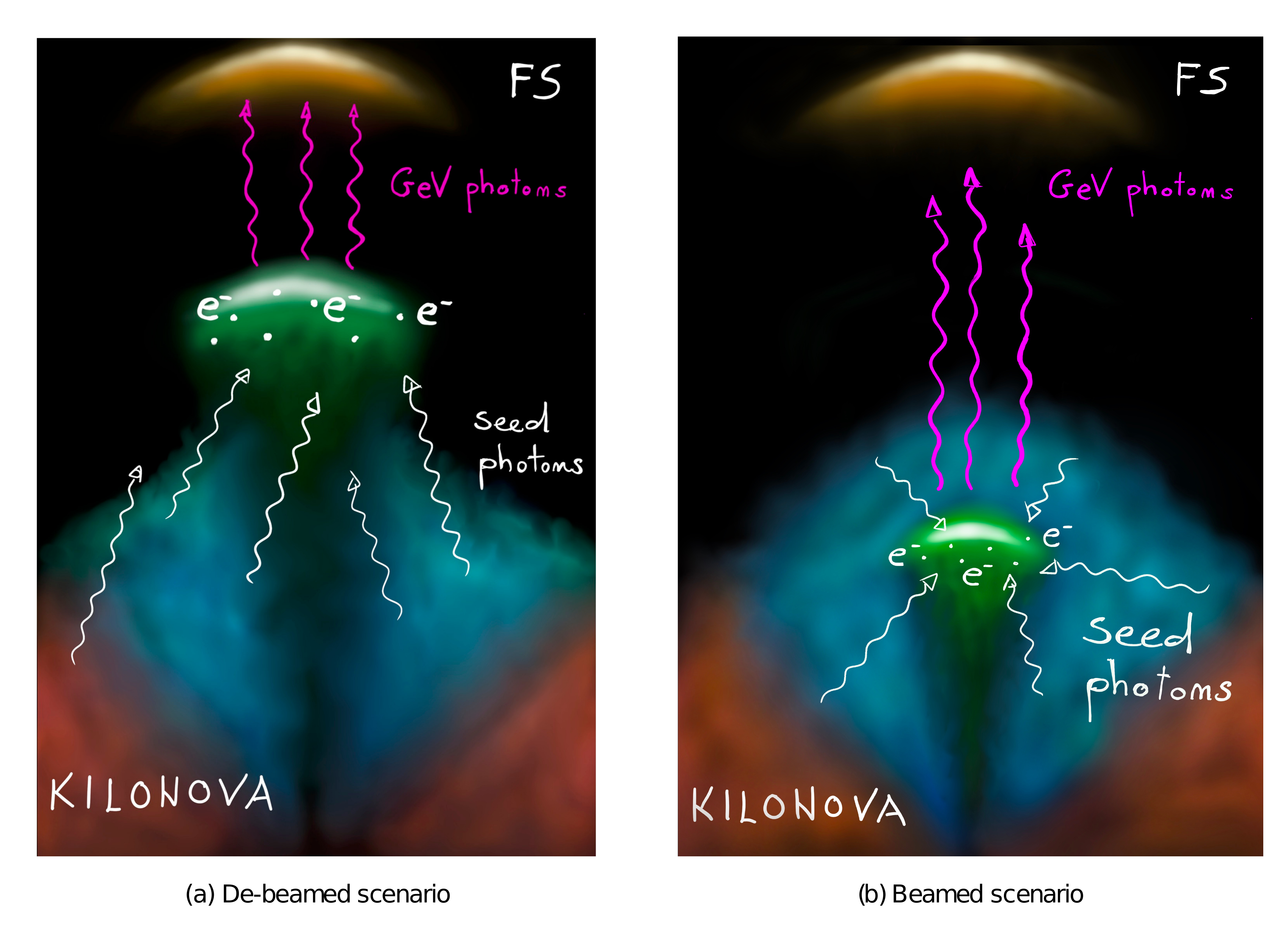} 
	\caption{}
    \label{fig:draw}
\end{figure*}\\

\clearpage

\bibliographystyle{naturemag}
\bibliography{references}

\section*{ Figure Legends}\label{sec:methods}

\bmhead{Fig. 1: \lat\ detection of GRB 211211A}
Test Statistic (TS) maps (a, b, c) centered at the GRB position (R.A. $=212.272^{\circ}$, Dec. $= 27.884^{\circ}$) and the GRB light curve and spectrum (d). With respect to the GRB 211211A trigger time $t_0$, the TS maps are shown: (a) one day before $t_0$, (b) during the first 6 hours after $t_0$, and (c) one day after $t_0$. While a significant excess has been observed within 1-20\,ks after the GRB trigger time reaching TS of $\sim43$, no significant excess has been observed in any other time-bins. The bottom panel (d) shows the GRB 211211A light curve between 0.1-1\,GeV. Flux upper limits are shown (in black) for the epochs with $\rm TS<9$. The two spectra shown in (d, right panel) are between 
4-7\,ks (blue) and 12-20\,ks (red).
We report error bars and upper limits with 1 and 2$\sigma$ confidence level, respectively.

\bmhead{Fig. 2: High energy light curves of GRBs observed by \lat}
Long (in green) and short (in blue) LAT light curves from the 2nd \lat\ GRB catalog \cite{ajello2019} compared to GRB 211211A emission (in brown).\\
We show the LAT flux in the energy band of 0.1-10 GeV as a function of time from the burst trigger time $\rm T_{0}$ obtained through time-resolved analysis with a test statistic $\rm TS > 9$ (a).
For the sub-sample of long and short GRBs with redshift estimates ($\sim34$ sources), we show the isotropic-equivalent LAT luminosity $\rm L_{iso}$ (b) as a function of the rest-frame time after trigger. Error bars are reported with 1$\sigma$ confidence level.

\bmhead{Fig. 3: Multi-wavelength light curves and spectra of GRB 211211A}
The figure shows light curves (a) and spectral energy density distributions (SEDs, b) of GRB~211211A. Fluxes inferred from observations are shown by circles, squares or stars with error bars, while downward-pointing triangles indicate three-sigma upper limits (multiplied by $10^3$ in the case of radio observations). In panel (b), butterfly-shaped symbols show one-sigma flux confidence regions for \xrt\ and \lat. Solid lines show our best-fitting model, consisting of emission from the forward shock and a kilonova. The model radio light curves are multiplied by $10^3$, consistently with the corresponding datapoints. Dotted lines single out the kilonova contribution. The SEDs are relative to the time bins marked with vertical grey bands in (a).

\bmhead{Fig. 4: external inverse Compton model contribution}
Same as Fig. \ref{fig:grb1_sed_lc}, but with dashed lines showing the contribution from kilonova photons up-scattered by external inverse Compton of relativistic electrons within a late-time, low-power jet, possibly sustained by fallback accretion on the merger remnant (see Methods).

\bmhead{Fig. 5: The interaction between the low-power jet and the kilonova}
Artistic impression of the scenario explaining the {\it Fermi}-LAT observations of the GRB 211211A. Seeds photons emitted from the kilonova ejecta (in red and blue) are scattered via inverse Compton by electrons in a low-power jet (in green) launched at late times. The External Inverse Compton can occur at a radius above (a) or below (b) the kilonova surface. The external forward shock of the relativistic jet giving rise to multi-wavelength afterglow emission is shown in yellow.

\section*{Methods}\label{sec:methods}
\bmhead{{\textit Swift}/XRT} 
We have downloaded the data provided by the X-Ray Telescope (0.3 - 10 keV, XRT) on-board the Neil Gehrels Swift Observatory ({\it Swift}) from the {\it Swift} Science Data Center supported by the University of Leicester \citesupp{Evans2009}. Eight time-bins (3.5 - $140$ ks from the GRB trigger time) in the Photon Counting mode were selected for the time-resolved spectral analysis to evaluate the temporal and spectral evolution of the X-ray emission during the afterglow phase. We have performed spectral analysis of XRT time-resolved spectra with XSPEC (v12.12.0) using the CSTAT likelihood. We have fitted the 0.3 - 10 keV spectra by a simple power law taking into account the Galactic absorption by applying the multiplicative \texttt{tbabs} model. Galactic absorption by neutral hydrogen in the direction of the GRB is estimated from \citesupp{Kalberla2005}. We have additionally fitted all the time-resolved spectra by applying also \texttt{ztbabs} model to account for the intrinsic absorption. We have found the intrinsic column density of the neutral hydrogen being consistent with zero. Therefore, we have further excluded the intrinsic absorption from our modelling. The best fit parameters, i.e. unabsorbed flux and the photon index were used for the modeling of the afterglow emission. 

\bmhead{{\textit Swift}/UVOT}
The \textit{Swift}/UVOT carried out observations of GRB 211211A between $\rm T_0+92$ s and $\rm T_0+3.3$ d. The GRB optical and UV afterglow has been detected in all UVOT filters until $\rm T_0+1.3$ d \citesupp{UVOT_GCN_Circular,Rastinejad2022}. We retrieved and analysed the \textit{Swift}/UVOT data obtained with the white filter from the Swift archive\footnote{https://heasarc.gsfc.nasa.gov/cgi-bin/W3Browse/swift.pl}. The afterglow magnitudes have been obtained with the task \texttt{uvotsource}, part of the \textsc{HEASoft} software package\footnote{https://heasarc.gsfc.nasa.gov/docs/software/heasoft/}, using a circular extraction aperture of $3''$ radius (in order to minimise contamination from the nearby host galaxy) and a background circular region of $20''$ radius. An aperture correction has been applied to report the obtained magnitudes to the standard $5''$ aperture. For the afterglow light curves in the other UVOT filters we refer to the magnitudes values reported in \citesupp{Rastinejad2022}.\\
While we did not include, as a conservative choice, the white filter data in the model fitting, we estimated \textit{a posteriori} its compatibility with our modelling as follows. Using the appropriate transmission curve\footnote{\url{https://www.mssl.ucl.ac.uk/www_astro/uvot/uvot_instrument/}} we found that, assuming a spectrum with an intrinsic power law flux density $\rm F_{\nu}\propto\nu^{-0.65}$ (corresponding to the spectral shape predicted by our model) and accounting for Galactic interstellar dust absorption with $E(B-V)=0.015$ \citesupp{Schlafly2011}, the extinction-corrected white filter AB magnitude can be transformed into the equivalent $u$-filter one by a $+0.068$ magnitude correction. The earliest $u$-filter data point in Fig.~\ref{fig:grb1_sed_lc}, obtained in this way, is in excellent agreement with the model prediction and it shows that the optical afterglow flux density was most likely rising at times $t\lesssim 5000\,\mathrm{s}$.

\bmhead{{\textit XMM}-Newton}
We obtained two epochs of observations of the field of GRB\,211211A with \textit{XMM-Newton} at mid-times of $\sim 9.6$ and $\sim 50.9$ days after the burst, lasting 40 and 67 ks (EPIC/pn exposure), respectively. We relied on data products released through the Processing Pipeline Subsystem (PPS), with standard filtering for the background flares, resulting in 12.7 and 38.3 ks effective exposure time, respectively. Fully consistent results were obtained from an independent custom reduction carried out with the \textit{XMM-Newton} Science Analysis Software (\textsc{SAS}). No clear X-ray source is detected at the afterglow position. From the resulting EPIC/pn images we derive $3\sigma$ upper limits of $\sim 3.6 \times 10^{-3}$ and $\sim 1.5 \times 10^{-3}$ cts s$^{-1}$ for the first and second epoch, respectively. Assuming the spectral parameters derived by {\it Swift}/XRT\footnote{\url{https://www.swift.ac.uk/xrt_spectra/01088940/}}, i.e. Galactic $\rm N_H = 1.76 \times 10^{20}$ cm$^{-2}$ and photon index $\Gamma = 1.5$, these values translate into unabsorbed flux limits of $\sim 1.1 \times 10^{-14}$ and $\sim 4.8 \times 10^{-15}$ erg s$^{-1}$ cm$^{-2}$in the 0.3 - 10 keV energy range.\\ 
However, we report that running a targeted search at the GRB afterglow position on the first epoch of EPIC/pn data through the \textsc{Sosta} (\textsc{SOurce STAtistcs}) tool\footnote{part of the \textsc{HEASoft/Ximage} software} a possible excess is detected with a count rate of $(3.2\pm1.0)\times 10^{-3}$ cts s$^{-1}$. Assuming the above spectral parameters, this translates into an unabsorbed 0.3-10 keV flux of $(1.02\pm 0.32) \times 10^{-14}$ erg s$^{-1}$ cm$^{-2}$ . However, given the low-significance, it is not possible to assess if this excess is due to a real source or just to a spurious fluctuation. Considering this data point as an upper limit or a detection has no consequences for the conclusions of this paper.

\bmhead{Telescopio Nazionale Galileo (TNG)}
Near-infrared (NIR) observations of GRB\,211211A were carried out with the Italian 3.6-m TNG telescope, sited in Canary Island, using the NICS instrument in imaging mode. A series of images were obtained with the H filter on 2021-12-16 from 05:51:36 UT to 07:00:51 UT (i.e. at a mid time of about 4.7 days after the burst). The image reduction was carried out using the {\it jitter} task of the ESO-eclipse package\footnote{https://www.eso.org/sci/software/eclipse/}.  Astrometry was performed using the 2MASS\footnote{https://irsa.ipac.caltech.edu/Missions/2mass.html}
catalogue. No source is detected at the optical and NIR counterpart position down to a $3 \sigma$ upper limit of H $> 20.5$ mag (Vega) or H $> 21.9$ mag (AB). 

\bmhead{Optical/NIR data}
GRB 211211A has been followed-up by numerous optical telescopes. We have selected observations in several bands (AB system) from the GCN Circulars Archive for the afterglow modelling. We include in our analysis r-band data from the NEXT-0.6m optical telescope \citesupp{GCN.31213}, Nordic Optical Telescope (NOT, \citesupp{GCN.31221}), MITSuME \citesupp{GCN.31217}, Himalayan Chandra Telescope (HCT, \citesupp{GCN.31227}), LCO 1-m Sinistro instrument \citesupp{GCN.31214}, Devasthal Optical Telescope \citesupp{GCN.31299}, Zeiss-1000 telescope of SAO RAS \citesupp{GCN.31234}; i-band data from CAFOS/2.2m CAHA \citesupp{GCN.31218, GCN.31228}, MITSuME \citesupp{GCN.31217} and k-band data from Gemini-North Telescope \citesupp{GCN.31235}. We show in Fig.~\ref{fig:grb1_sed_lc} (green star at $\rm \sim 4000 s$) also the r-band flux density derived by converting a KAIT white filter observation\citesupp{GCN.31203} following Ref.\citesupp{Li2003_KAIT}, but we conservatively do not include it in the modelling.

\bmhead{Very Large Array}
Observations with the Karl G. Jansky Very Large Array (VLA) were performed 35 (January 15, 2022), 39 (January 19, 2022) and 77 (February 26, 2022) days post-burst (PI: Giarratana; project code: 21B-370) at the central frequencies of 3 (S-band), 6 (C-band) and 10\,GHz (X-band), with a bandwidth of 2, 4 and 4\,GHz, respectively. The distance between the target and the phase calibrator (J1407$+$2827) is about 0.7$^\circ$. Each observation
started with scans of the flux and bandpass calibrator (J1331$+$3030). The data were calibrated using the custom \textsc{casa} pipeline (Version 6.2.1, \citesupp{McMullin2007}) and visually inspected for possible radio frequency interference. The final images were produced with the \texttt{tclean} task in \textsc{casa} (Version 5.1.0.). We did not detect any point-like transient consistent with the {\it Swift}/XRT position of the burst.\\

\bmhead{\textit{Fermi}/LAT}
The Large Area Telescope (LAT) on board \textit{Fermi} is sensitive to the gamma-ray photons in the energy band 0.1-300\,GeV \citesupp{2013ApJS..209...34A}. We use \textsc{gtburst} tool to extract and analyse the data. We define a region of interest (ROI) of 10$^{\circ}$ centred at the burst position (R.A. $=212.272^{\circ}$, Dec. $= 27.884^{\circ}$) provided by {\it Swift}/BAT. 
We initially perform a standard LAT unbinned likelihood analysis. The null hypothesis is given by the baseline likelihood model which includes the isotropic particle background (\texttt{isotr. template} in \textsc{gtburst}), galactic and extra-galactic high energy components from the Fermi 4th Catalog (4FGL, \citesupp{2020ApJS..247...33A}) with fixed normalisation (\texttt{template (fixed norm.)}). If a new source is present in the field of view (FoV), its addition to the model should describe the observed data better, given a certain position and spectral model. To assess if the improvement is significant, we use the Likelihood Ratio Test (LRT), as described in \citesupp{2019ApJ...878...52A}. For each time-bin, the LRT returns best-fit spectral values, as well as the relative test statistic   ${\rm TS} = -2 \times \ln \left( L_\mathrm{max,0}/L_\mathrm{max,1}\right)$, 
where $L_{\rm max,0}$ is the maximum likelihood for the null hypothesis (background only), and L$_{\rm max,1}$ is the maximum likelihood for a model with the additional source at a given location and with a given spectral shape. In this analysis, we use the GRB position estimated by \bat\ and a power law spectral model (\texttt{powerlaw2} using \textsc{gtburst}).\\
The spectral parameters are only reported when the test statistic (TS) is larger than 9 in a given time-bin. The choice of the time-bins is driven by the visibility of the source in the FoV of the telescope satisfying the condition of zenith-angle below 100$^{\circ}$ and angular distance (source off-axis angle, $\theta$) from the center of the LAT FoV less than $60^{\circ}$. The values of zenith angle of 100$^{\circ}$ are chosen in order to minimise the effect of the Earth occultation.\\
We find only two time-bins, 4-7\,ks and 12-20\,s after the trigger time ($\rm t_0 = 660921004\, s$ Mission Elapsed Time - MET) with $\rm TS\geq9$, while in the other time bins we obtain $\rm TS<9$, resulting in flux upper limits.\\
We choose the LAT data, in order to match the temporal bins covered by \xrt. We select five epochs: 0.90-1.34 ks, 4-7 ks, 12-20 ks, 20-110 ks, and 110-610 ks. We find that $\rm TS\geq 9$ only between 4-7 ks and 12-20 ks, with TS of 9 and 31 corresponding to $\sim3\sigma$ and $\sim5\sigma$ detections, respectively. For the first epoch, we estimate a flux $ F_{\rm LAT,1}$ = $(5.8 \pm 4.0) \times 10^{-10}\ \rm erg\ s^{-1}\ cm^{-2}$
(0.1-1 GeV) and photon index $ \Gamma_{\rm LAT,1}$ =  $-2.8 \pm 0.7$, 
while for the second epoch $F_{\rm LAT,2}$ = $(5.1 \pm 2.0) \times 10^{-10} \rm erg\ s^{-1}\ cm^{-2}$ and photon index $\Gamma_{\rm LAT,2}$ = -3.1 $\pm$ 0.6 (see Fig.~\ref{fig:TSmaps}). We detected 9 photons from the GRB during the first 20\,ks of observation with the probability of association with the GRB (p; estimated with \textsc{gtsrcprob} and P8R3\_TRANSIENT010E\_V2 as instrument response function) larger than 0.9. 
The standard criteria for detecting a GRB according to \citesupp{2019ApJ...878...52A} is to have more than 3 photons with p$>$0.9. The highest energy photon has been detected at 13\,ks (at a position 0.32$^{\circ}$ away from the GRB location and with an associated probability of p=0.97). This photon has an energy of 1.74\,GeV. The
properties of the photons, such as energy, GRB association probability, distance from the source and the arrival time from the trigger time are reported in Extended Data Table \ref{tab:GeVPhotons}.

\bmhead{Finding the location of the GeV excess}
We test the presence of a GeV source with the help of a test statistics  map (TS map). In TS map, we divide the ROI (12$^{\circ}$ around the GRB position) in several pixels with a side of 0.8$^{\circ}$. We perform the same LAT likelihood analysis described in the previous section. We fix all spectral parameters, including the Galactic and isotropic diffuse emission templates \citesupp{abdalla2019}, except for the normalization factor of the spectra, which are left free to vary. The analysis returns a test statistic value for each pixel, assessing which are the positions in the ROI where the detected emission is coming from.\\
We applied this strategy to three time-bins: one day before, one day after and the day of the GRB 211211A trigger. Since this analysis explores a longer time duration than 200\,s, the event class \texttt{P8R3\_TRANSIENT010E\_V2} is used as mentioned in \citesupp{2019ApJ...878...52A}. We performed a LRT by considering the time intervals only when the border of
the ROI is at a zenith angle less than 100$^\circ$, which reduces the contamination from the Earth limb.\\
We show the results in Fig.~\ref{fig:TSmaps}. We obtain a maximum $\rm TS \simeq 43$ in the TS map made during the first 20 ks after the burst, resulting in a $>5\sigma$ detection, in spacial coincidence with the GRB, hence confirming the existence of the GeV source. Conversely, in the day before and after the burst, the maximum TS does not reach 9 in coincidence with the burst position.

\bmhead{Ruling out external high energy contamination}
The ROI of GRB 211211A includes 5 sources within $5^\circ$ from the burst location: 4FGL J1410.4+2820, 4FGL J1417.9+2543, 4FGL J1424.1+2917, 4FGL J1351.9+2847, and 4FGL J1350.8+3033, with distances 0.55$^{\circ}$, 2.93$^{\circ}$, 3.59$^{\circ}$, 3.87$^{\circ}$, and 4.79$^{\circ}$, respectively. Among these sources, 4FGL J1410.4+2820 is located at a distance of $0.5^\circ$, which is smaller than the LAT point spread function $68\%$ containment angle ($\sim 1^\circ$ at 1\,GeV \citesupp{ajello2021}).
This source is associated with RX J1410.4+2821 (R.A. $=212.623^{\circ}$, Dec. $=28.342^{\circ}$), a BL Lacertae object (BL Lac, \citesupp{massaro2016}).
In order to rule out the possibility that the GeV photons detected at the location of the GRB are associated with a flaring state of the BL Lac object and estimate the possible contamination in our detection by the BL Lac flux, we analyzed the data of the BL Lac object 4FGL J1410.4+2820 using the \textsc{enrico tools} \citesupp{2013ICRC...33.2784S}. We selected
\texttt{P8R3\_SOURCE\_V2} as the response function. We used events from
0.1 - 300\,GeV selected within a 10$^\circ$ ROI centered at 4FGL J1410.4+2820 and having a zenith distance below 100$^\circ$
to avoid contamination from the Earth’s limb. The diffuse Galactic and
isotropic components were modelled with the files \texttt{gll\_iem\_v07.fits}
and \texttt{iso\_P8R3\_SOURCE\_V2.txt}, respectively. We included the point sources in the 4FGL located in the 10$^\circ$ ROI and an additional surrounding 5$^\circ$-wide annulus. The spectral slopes  were fixed to their 4FGL values, while the normalization of the sources within the ROI as well as the diffuse components are kept free to vary. We analyzed monthly-binned data of the BL Lac object before the trigger time of the GRB (658293004-660921004\,s MET) and after (661007404-663635404\,s MET) excluding one day around the GRB burst (660921004-661007404\,s MET). The spectral analysis of the BL Lac for one month 
before and after the trigger returns upper limits of values $\rm 7 \times 10^{-13}\ erg\ s^{-1}\ cm^{-2}$  and $\rm 2 \times 10^{-12}\ erg\ s^{-1}\ cm^{-2}$, respectively. These upper limits are $\sim2$ orders of magnitude smaller than the flux obtained from the LAT excess in temporal coincidence with the GRB. We also included the time-averaged broadband spectrum of the BL Lac in 12 years of observation\footnote{\url{https://fermi.gsfc.nasa.gov/ssc/data/access/lat/12yr_catalog}} (see Extended Data Fig. \ref{fig:blazar_sed}, \citesupp{2022arXiv220111184F}) which is consistent with the flux upper limits derived from the monthly-binned data from the BL Lac object.

\bmhead{Comparisons with other LAT emissions from GRBs}
The first sGRB detected by \lat, GRB 090510 \citesupp[$\rm z=0.9$;][]{2010ApJ...716.1178A}, was observed in the GeV band during the first 1000\,s of observation and then the emission fades away with no detectable emission at later time (after 1\,ks).\\ 
GRBs emitting high energy radiation detected by \lat\ are collected in the second \lat\ GRB Catalog (2FLGC, \citesupp{ajello2019}). The catalog contains $\sim200$ sources starting from \textit{Fermi} launch up to the end of 2018, for a total of ten years. We compare the  GRB 211211A with the populations of long and short GRBs in the catalog. Our classification of long or short bursts is based on the $\rm T_{90}$ estimate provided by \gbm. We consider only emissions that reach a test statistics $\rm TS\geq9$ in the time-resolved analysis performed in \citesupp{ajello2019}.\\
By comparing the fluxes in the 0.1-10 GeV energy band (Fig.~\ref{fig:LAT_catalog}, panel a) we note that GRB 211211A observations lie in the tail of light curves of the long GRB population. If we consider the isotropic-equivalent LAT luminosity $\rm L_{iso}$ (Fig.~\ref{fig:LAT_catalog}, panel b) for a sub-sample of GRBs with redshift measurements, we observe that this emission is significantly fainter with respect to both short and long populations. This suggests that the LAT emission of GRB 211211A is intrinsically fainter, and it would be hardly detectable for sources placed at larger distances (i.e. $>$ 350 Mpc). \\
In Extended Data Fig.~\ref{fig:LAT_catalog2} (panel a)  we compare the prompt duration of the burst $\rm T_{90}$ with the time at which LAT started to detect HE emission from the GRB. The GRB 211211A occupies the upper-right corner of the plot, together with other long GRBs showing LAT late-time emission. 
In Extended Data Fig. \ref{fig:LAT_catalog2} (panel b), we compare flux in the energy range 0.1 - 10 GeV and spectral index obtained through a time-integrated analysis along all the LAT duration. We observe that GRB 211211A occupies a region of the plot relatively far from the clustered region occupied by long GRBs and the sparse distribution of short GRBs.\\

\bmhead{Search for similar GeV component in nearby GRBs}\label{sec:NOHE}
Four GRBs closer than 350\,Mpc have been detected by {\it Swift}/BAT and {\it Fermi}/GBM so far: GRB 111005A, GRB 100316D, GRB 171205A, and GRB 190829A with redshift of 0.013, 0.059, 0.037, and 0.078, respectively. The GRB 100316D, GRB 171205A, and GRB 190829A are typical long GRBs with associated SNe observed. The LAT data analysis for these three GRBs shows no detection from the source within one day from the trigger time. The corresponding upper limits are the following: 1.15$\times 10^{-10}$ erg\,s$^{-1}$cm$^{-2}$ , 1.35$\times 10^{-10}$ erg\,s$^{-1}$cm$^{-2}$  and 1.38$\times 10^{-10}$ erg\,s$^{-1}$cm$^{-2}$ , respectively. GRB 111005A is a long type GRB \citesupp{2018A&A...616A.169M} without an associated SN. The \lat\ observations for this GRB also result in flux upper limits of 7.9$\times 10^{-10}$ erg\,s$^{-1}$cm$^{-2}$ , 9$\times 10^{-10}$ erg\,s$^{-1}$cm$^{-2}$ , 1.8$\times 10^{-10}$ erg\,s$^{-1}$cm$^{-2}$ , 1.78$\times 10^{-10}$ erg\,s$^{-1}$cm$^{-2}$  and 1.74$\times 10^{-10}$ erg\,s$^{-1}$cm$^{-2}$  for the five consecutive days after the trigger (t$_0$ = 339494716.22  MET). The sGRB 170817A, associated with the BNS merger event GW170817, was not detected by LAT during the Gravitational Wave trigger as the telescope was transiting through the South Atlantic Anomaly, preventing to put constraints on the high energy component at the time of the merger. At later times (from minutes to days after the BNS merger), the LAT observations do not show any detection but enable to place flux upper bounds \citesupp{2018ApJ...861...85A}. In the time-bin between 1153-2027\,s after the GW trigger an upper limit of 4.5$\times 10^{-10}$ erg\,s$^{-1}$cm$^{-2}$ was estimated in the 0.1 - 1 GeV energy range corresponding to an equivalent isotropic luminosity upper limit of 9.7$\times 10^{43}$ erg\,s$^{-1}$. In the same time interval, we have a flux upper limit whose corresponding luminosity upper limit is two orders of magnitude brighter than the one measured for GW170817. Instead, in the time interval of our GeV detecion, Fig. 3 of \citesupp{2018ApJ...861...85A} shows a flux upper limit which is on the order of $10^{-9}$ erg\,s$^{-1}$cm$^{-2}$. By correcting for the distances, the flux of GRB 211211A for a source at the same distance as GW170817 would have been 
brighter than this upper limit. The GeV emission discovered for GRB 211211A was not present in GW170817. This can be explained within our scenario interpreting the GeV emission; GW170817 was observed off-axis (a viewing angle of about 15 deg from the jet) where the EIC emission is expected to be severely suppressed due to relativistic beaming. In addition, the very dense kilonova ejecta could absorb most of the off-axis GeV emission via Bethe-Heitler process, further preventing its detection for off-axis observers.

\bmhead{Forward shock model}
The GRB emission is thought to be produced within a collimated, relativistic outflow (i.e.\ a jet) that moves in a direction close to the line of sight \citesupp[e.g.][]{Kumar2015}. As the jet expands into the interstellar medium (ISM) at  above the local sound speed, a forward shock (FS) forms and propagates in the ISM \citesupp[e.g.][]{Sari1998,Panaitescu2000}. 
We model the dynamics and emission of the FS, assuming the line of sight to be on the jet axis and the ISM to be homogeneous with number density $n$, following  \citesupp{Salafia2022_GRB190829A}, with the only minor difference that we model the lateral expansion of the shocked region following the recipe from \citesupp{Granot2012} instead of assuming sound-speed expansion as in \citesupp{Salafia2022_GRB190829A}. While the difference in the light curves obtained with the two methods is very small, we found the sound-speed model to produce slight discontinuities in the derivative of some light curves, which we believe is an artifact. The emission model includes both synchrotron and synchrotron-self-Compton (SSC) from electrons accelerated at the FS, but we find that, for our best-fit parameters, the SSC contribution is negligible in the relevant bands. In the model, a fraction $\chi_\mathrm{e}$ of the ISM electrons that cross the shock is assumed to be accelerated to relativistic speeds. Accelerated electrons are assumed to be injected into the shocked fluid with a Lorentz factor $\gamma$ distribution $\mathrm{d}n/\mathrm{d}
\gamma\propto \gamma^{-p}$ for $\gamma\geq \gamma_\mathrm{m}$, to hold a fraction $\epsilon_\mathrm{e}$ of the shocked fluid energy density and to be subject to cooling due to synchrotron and SSC emission (the cooling due to the latter is computed including Klein-Nishina effects). Small-scale turbulence is assumed to produce an effectively isotropic magnetic field which holds a constant fraction $\epsilon_\mathrm{B}$ of the shocked fluid energy density. The model is entirely determined by 8 parameters, namely the jet isotropic-equivalent kinetic energy $E$, its bulk Lorentz factor $\Gamma$ and half-opening angle $\theta_\mathrm{j}$, the ISM number density $n$, and the `microphysical' parameters $\epsilon_\mathrm{e}$, $\epsilon_\mathrm{B}$, $\chi_\mathrm{e}$ and $p$. 

\bmhead{Kilonova model}
During and after the merger of a binary of two compact objects including at least one neutron star (NS), outflows of neutron-rich material can be produced by a variety of mechanisms. Ejecta that remain cold (i.e.\ are not heated by shocks) and do not receive intense neutrino irradiation retain an extremely low proton fraction $Y_\mathrm{e}=n_\mathrm{p}/(n_\mathrm{n}+n_\mathrm{p})\lesssim 0.05$ (e.g.\ \citesupp{Hotokezaka2013}, where $n_\mathrm{p}$ and $n_\mathrm{n}$ are the number densities of protons and neutrons, respectively). Tidal forces (especially acting on the least massive component of the binary) can unbind cold material, mainly along the orbital plane and over ms timescales, with high speeds ($0.1-0.3$ {\it c}), potentially large masses (up to 0.1 M$_\odot)$, and low $Y_\mathrm{e}$ \citesupp{Hotokezaka2013,Radice2018}; in the case of a binary NS (BNS) merger, shocks at the colliding surfaces of the two NS can eject low amounts ($\lesssim 10^{-3}$ M$_\odot$) of comparably high-speed material, with high $Y_\mathrm{e}$, preferentially along the polar direction; again in the BNS case, if the merger does not lead to a prompt collapse to a black hole (BH), violent oscillations and/or bar modes in the immediate post-merger remnant can produce  copious mass ejection (few$\times 10^{-2}$ M$_\odot$) of neutrino-irradiated material with relatively high $Y_\mathrm{e}$ \citesupp{nedora2019}; last, but not least, winds driven by neutrino energy deposition (leading to high $Y_\mathrm{e}$) and more prominently by viscous angular momentum transport (leading to a broad range of $Y_\mathrm{e}$ values) in the accretion disk around the merger remnant can unbind a large fraction ($\sim 10-30\%$) of the disk mass, with relatively low speeds $\lesssim 0.1$ \citesupp[e.g.][]{Just2015,Siegel2018}.\\
During the first stages of expansion, nucleosynthesis by rapid neutron capture (r-process) takes place within these outflows. If $Y_\mathrm{e}$ is sufficiently low (about $Y_\mathrm{e}\lesssim 0.2$, e.g.\ \citesupp{Lippuner2015}) heavy r-process elements (Lanthanides and Actinides) can be produced. Their complex valence electron structure results in an extremely high opacity to photons in the infrared-to-ultraviolet wavelength range \citesupp{Kasen2013}. As a result, outflows that start off with $Y_\mathrm{e}\lesssim 0.2$ are expected to have extremely high opacities $\kappa \gtrsim 10\,\mathrm{cm^2\,g^{-1}}$ and thus long diffusion times. Lanthanide-free outflows (those initially with $Y_\mathrm{e}\gtrsim 0.2$) are instead expected to have lower opacities $0.5\lesssim \kappa/\mathrm{cm^2\,g^{-1}} \lesssim 3$. Regardless of the presence of Lanthanides, the radioactive decay (mainly beta) of unstable isotopes in these outflows is thought to constitute a heating source for the ejecta with a robust heating rate that depends very weakly on the exact composition \citesupp{Korobkin2012}. Such heating is the power source of the so-called `kilonova' (KN) emission that is produced as these outflows expand \citesupp{Li1998,Metzger2019}.\\
In the modelling behind this work, we did not attempt at capturing the complexity of the kilonova emission in presence of several outflows with different masses, speeds, opacities and geometries, given the insufficient detail in the available data. We instead opted for adopting the simple, semi-analytical, one-component, isotropic model from \citesupp{grossman2014}. This model is entirely specified by three parameters, namely the ejecta mass $m_\mathrm{ej}$, maximum speed $v_\mathrm{max,ej}$ and constant grey opacity $\kappa$.\\
The inferred kilonova ejecta mass is $ m_\mathrm{ej} = 2.0_{-0.6}^{+0.9}\times 10^{-2}\,\rm M_{\odot}$, expanding with an average velocity of $v_\mathrm{ej} \sim 0.5\ v_\mathrm{max,ej} = 0.10_{-0.04}^{+0.07}\ c$ \citesupp{grossman2014, Wollaeger2018} and with a relatively low opacity $\kappa = 0.6_{-0.3}^{+0.8} \,\rm cm^{2}\ g^{-1}$. These properties are compatible with winds from a hyper-massive proto-neutron star remnant (HMNS, \citesupp{ciolfi2020}). Alternatively, such ejecta properties are in general agreement with those expected from material ejected due to enhanced angular momentum transport in the highly rotating, oblate remnant due to the formation of spiral arms (the `spiral wave wind' described in \citesupp{nedora2019}). Both interpretations favour the hypothesis that the merger remnant passed through a HMNS phase before collapsing. A black hole-neutron star scenario seems less favoured with respect to BNS merger, especially due to the velocity of the low-opacity component which, in a black-hole neutron star merger, would have to be produced in the form of winds from the accretion disk around the merger remnant, but with a substantially lower expected velocity $v_\mathrm{ej}\sim 0.03-0.06\,c$ \citesupp{fernandez2020}).\\

\bmhead{Model fitting}

We fit our 11-parameter model (8 parameters for the FS and 3 for the KN) to the available light curve data as follows. Let us define our observations as a set of flux densities $\lbrace F_{\nu,i}\rbrace_{i=1}^{N_\mathrm{F_\nu}}$ measured at radio, near-infrared, optical and ultraviolet frequencies at times $t_i$, and a set of fluxes and photon indices $\lbrace F_i,\alpha_i\rbrace_{i=1}^{N_F}$ measured in X- and $\gamma$-rays. Each flux density measurement contributes an additive term to our log-likelihood, which reads, in the case of detections,
\begin{equation}
    \log\mathcal{L}_{F_\nu,i} = -\frac{1}{2}\frac{(F_\mathrm{\nu,m,i}(t_i) - F_\mathrm{\nu,i})^2}{\sigma_i^2} - \frac{1}{2}\ln\left(2\pi\sigma_i^2\right),
    \label{eq:Fnu_likelihood}
\end{equation}
where $F_\mathrm{\nu,m,i}(t_i)$ is the flux density predicted by the model at the corresponding time and frequency, and $\sigma_i=\sqrt{\sigma_{\mathrm{obs},i}^2+f_\mathrm{sys}^2F_\mathrm{\nu,m,i}^2}$ is the assumed uncertainty, which is the sum square of the formal uncertainty $\sigma_{\mathrm{obs},i}$ associated to the observation and an unknown systematic contribution to the error, parametrized by the dimensionless constant $f_\mathrm{sys}$ (which therefore constitutes an additional model parameter). The normalization term (the second term on the right-hand side of Eq.~\ref{eq:Fnu_likelihood}) is included as it effectively represents a penalty for higher values of $f_\mathrm{sys}$. In the case of upper limits, our log-likelihood term becomes a simple one-sided Gaussian penalty, namely
\begin{equation}
    \log\mathcal{L}_{\mathrm{UL},i} = -\frac{1}{2}\frac{\left\lbrace\max\left[(F_\mathrm{\nu,m,i}(t_i) - F_\mathrm{\nu,i}), 0\right]\right\rbrace^2}{(0.01 F_\mathrm{\nu,i})^2}.
    \label{eq:uplim_likelihood}
\end{equation}
For X and $\gamma$-ray fluxes, the additive term is
\begin{equation}
    \log\mathcal{L}_{F,i} = -\frac{1}{2}\frac{(F_\mathrm{m,i}(t_i) - F_\mathrm{i})^2}{\sigma_i^2} - \frac{1}{2}\ln\left(2\pi\sigma_i^2\right) - \frac{1}{2}\frac{(\alpha_\mathrm{m,i}(t_i) - \alpha_\mathrm{i})^2}{\sigma_{\alpha,i}^2},
    \label{eq:F_likelihood}
\end{equation}
where again $\sigma_i=\sqrt{\sigma_{\mathrm{obs},i}^2+f_\mathrm{sys}^2F_\mathrm{m,i}^2}$ as for the flux densities,  $\sigma_{\alpha,i}$ is the uncertainty on the observed photon index (we assume no systematic uncertainty on the photon index), and the model photon index $\alpha_\mathrm{m,i}$ is simply defined as the average of the slope of the model photon spectrum over the instrument band.\\
Adopting log-uniform priors on all parameters except $p$ (for which we use a uniform prior), within the bounds given in  Extended Data Table~\ref{tab:model_fitting}, we sampled the posterior probability density with a Markov Chain Monte Carlo (MCMC) approach using the \texttt{emcee} python package \citesupp{ForemanMackey2013}, employing $N_\mathrm{walk} = 4\times N_\mathrm{dim}=48$ walkers (where $N_\mathrm{dim}=12$ is the dimension of the parameter space). We initialized the walkers in a small $N_\mathrm{dim}$-dimensional ball around a point in our parameter space representing ``standard'' GRB afterglow parameters, $x=(52.5,0,2,-1,-1,-3,0,2.1,-2,0,-1,-1)$ where $x$ represents the parameters as listed in Extended Data Table \ref{tab:model_fitting}, and we performed $N_\mathrm{iter}=10000$  iterations, for a total of $N_\mathrm{iter}\times N_\mathrm{walk}=480000$ samples. The final mean auto-correlation time is $\sim 600$, and the posterior looks reasonably smooth and single-peaked. As a cross-check, we also run several shorter ($N_\mathrm{iter}\sim 2000$) chains with different starting parameters, and these all converged to the same parameter space region after a burn in. A corner plot demonstrating the features of our posterior is shown in Extended Data Fig.~\ref{fig:afterglow_model_corner_plot}, while summary information on the parameter credible ranges from the marginalised posteriors is reported in Extended Data Table~\ref{tab:model_fitting}.  Most parameters are relatively well constrained, except for the ISM density, whose posterior rails against the prior bound and it can therefore be only constrained to be $n/\mathrm{cm^{-3}}<10^{-4.2}$ ($95\%$ credible level, consistent with the large offset from the host galaxy and the absence of local absorption), and the fraction of accelerated electrons, that can only be constrained to be $\chi_\mathrm{e}<0.3$ ($95\%$ credible level, in agreement with expectations from particle-in-cell simulations, e.g. \citesupp{Spitkovsky2008}).

\bmhead{External Inverse Compton}

To explain the high energy (100 MeV - 1 GeV) excess at $\rm 10^{4}$ s found for GRB 211211A with respect to the standard synchrotron and SSC model of the afterglow, we invoke emission by External Inverse Compton (EIC). EIC scattering of soft seed photons by hot electrons in relativistic jets has been considered for a long time to explain the high energy emission in blazars \citesupp{Dermer1993}. 
Signatures of the EIC cooling of electrons in the GRB afterglow phase have been proposed by different authors. 
Prompt emission photons can be Comptonised in the reverse \citesupp{Beloborodov2005} and forward shock \citesupp{Fan2005} of the blast wave. The EIC radiation from upscattered prompt emission \citesupp{murase2010,murase2011}, X-ray \citesupp{Wang2006a,fan2008,Zhang2021b,murase2018} or UV flares \citesupp{Fan2006}, a dense ambient infrared photon field \citesupp{Zhang2020}, by the forward shock accelerated electrons can give a rise of the GeV radiation. The photons from the supernova shock break out \citesupp{Wang2006a} or cocoon \citesupp{Toma2009,Kumar2014,DeColle2018,Kimura2019,Zhang2021b} are also considered as seed photons for the EIC in the forward shock site or in the internal dissipation site, including also late-time dissipation related to the X-ray plateau emission (see \citesupp[][]{zhang2018book} for a general overview). \\
Given that the \lat\ spectrum is soft and the afterglow spectrum at lower energies (X--ray) is rising in $\nu F_{\nu}$ (see Fig. \ref{fig:grb1_sed_lc},  panel b), the EIC spectral component should be preferably narrow, favouring thermal seed photons. As late as $\rm 10^{4}$ s after the GRB, the rise of the kilonova makes its photons the natural and viable candidate seed photon source. \\ 
We first consider EIC scattering of the kilonova photons by hot electrons produced in the forward shock. To do so, we estimate the size $R_\mathrm{FS}$, bulk Lorentz factor $\Gamma_\mathrm{FS}$ and the typical electron Lorentz factor $\gamma_\mathrm{m,FS}$ at the forward shock, at time $T$ after the GRB, using the parameters close to the best fit parameters of the afterglow model: \begin{equation}
 \Gamma_\mathrm{FS} = 58 \, E_{53}^{1/8}\ n_{-4}^{-1/8}\ T_{4}^{-3/8},
 \end{equation}
 \begin{equation}
 R_\mathrm{FS} = 4 \times 10^{18} \,  E_{53}^{1/4}\ n_{-4}^{-1/4}\ T_{4}^{1/4} \, \rm cm,
\end{equation}
\begin{equation}
 \gamma_\mathrm{m,FS} = 8\times 10^3 \, \epsilon_\mathrm{e,-1.5}\ \chi_\mathrm{e,-1}^{-1}\ E_{53}^{1/8}\ n_{-4}^{-1/8}\ T_{4}^{-3/8},
\end{equation}
where $Q_x$ stands for $Q/10^x$ in cgs units, $Q$ being any of the model parameters. 
\\
Given the large radius of the forward shock, most of the seed photons are received from behind, therefore the seed photons energy density in the comoving frame of the forward shock region can be approximated as
\begin{equation}
 U_\mathrm{seed}^{\prime} = \frac{L_\mathrm{seed}}{4 \pi R_\mathrm{FS}^{2} \Gamma_\mathrm{FS}^{2} c}.
\end{equation}\\
The Lorentz factor of electrons cooling via the EIC on a dynamical time scale of $R_{\mathrm{FS}}/(\Gamma_{\mathrm{FS}} c)$ is 
\begin{equation}
 \gamma_{\rm c,EIC} = \frac{3 \pi m_{\rm e} c^{3}  R_{\mathrm FS} \Gamma_{\mathrm{FS}}^{3}} {\sigma_{\rm T} L_{\rm seed}} \approx 10^{13} E_{53}^{5/8}\ n_{-4}^{-5/8}\ T_{4}^{-7/8}
\end{equation}
clearly indicating that efficient extraction of energy from forward shock accelerated electrons through the EIC process is impossible: the kilonova luminosity of $ L_\mathrm{seed} \approx \rm  10^{40} \, erg/s$ is too low to account for the observed $\sim100$ MeV component with $L_\mathrm{excess}\rm \sim 5 \times 10^{45} \, erg/s$. If there were any other source of NIR/optical seed photons with the required luminosity, at $\rm 10^{4}$ s, it would overshine the observed optical afterglow emission.\\
Therefore, to account for the high energy emission by the EIC, we need to invoke a source of hot electrons at much smaller radii. \\ 
As a heuristic explanation, we assume a low-power jet to be present at the relevant late times (see Fig.~\ref{fig:draw}). This is not novel in GRBs: many long and short GRBs are followed by late-time X-ray flares \citesupp{Burrows2005,Falcone2006,Chincarini2007} and plateau emission \citesupp{Nousek2006,Liang2007}. These emission components are widely thought \citesupp{Zhang2006,Ghisellini2007} to be linked to late-time internal dissipation in a long-lived jet, which in compact binary mergers can be produced either by a highly-magnetised and fast-spinning proto-magnetar remnant \citesupp{Dai1998,Zhang2001} or by fallback accretion \citesupp{Rosswog2007}.\\
Therefore, we assume the presence of a source of hot electrons in the jet, which we call the `dissipation site', in the vicinity of the kilonova ejecta. In this scenario, we can constrain the parameters of the dissipation site by the following requirements: (1) the seed photons for the EIC scattering are the KN photons, 
(2) the dissipation in the low-power jet should not over-shine the observed optical and the X-ray afterglow emission by its synchrotron and SSC radiation.\\
We define $L_{\mathrm j}$, $\Gamma_{\mathrm j}$ and $R_\mathrm{j}$ as the luminosity, bulk Lorentz factor and dissipation radius of the low-power jet, respectively. $\chi_{e} = 0.1 \chi_\mathrm{e,-1}$ is the fraction of jet electrons accelerated at the jet dissipation site, with an energy distribution of $\mathrm{d}N_\mathrm{e}/\mathrm{d}\gamma\propto \gamma^{-p}$ and a minimum Lorentz factor, assuming $p=2.5$, of 
\begin{equation}
    \gamma_\mathrm{m} \approx 610 \frac{\epsilon_\mathrm{e,-1}}{\chi_\mathrm{e,-1}} (\Gamma^{\prime}-1),
\end{equation}
where $\Gamma^{\prime}$ is the relative bulk LF of one shell relative to another (assuming internal-shock-driven dissipation of jet kinetic energy).\\
Relativistic electrons in the jet lose their energy via synchrotron radiation, SSC and EIC on a time-scale (in the jet comoving frame)
\begin{equation}
    t_{\rm c}^{\prime} = \frac{\gamma m_\mathrm{e} c^2}{P_{\rm \rm syn}^{\prime} + P_{\rm SSC}^{\prime} + P_{\rm EIC}^{\prime}} = \frac{3 m_\mathrm{e} c}{4 \sigma_\mathrm{T} \gamma [U_{B}^{\prime}+U_{\rm syn}^{\prime}+U_{\rm ext}^{\prime}]},
\end{equation}
where $U_{B}^{\prime}$, $U_{\rm syn}^{\prime}$ and $U_{\rm ext}^{\prime}$ are the magnetic, synchrotron photon and external photon energy densities in the comoving reference frame, respectively, and we are assuming both SSC and EIC happen in the Thomson regime. 
The Lorentz factor of electrons which cool efficiently in a dynamical time-scale $t^\prime_\mathrm{dyn} = R_\mathrm{j}/(c\Gamma_\mathrm{j})$ is given by
\begin{equation}
    \gamma_{\rm c} = \frac{3 m_{\rm e} c^{2} \Gamma_{\rm j}}{4 \sigma_{\rm T} R_{\rm j}} [U_{B}^{\prime}+U_{\rm syn}^{\prime}+U_{\rm ext}^{\prime}] = \frac{\gamma_{\rm c}^{\rm syn}}{1+\frac{U_{\rm syn}^{\prime}}{U_{B}^{\prime}} + \frac{U_{\rm ext}^{\prime}}{U_{B}^{\prime}}}, 
\end{equation}
where $\gamma_{\rm c}^{\rm syn} = (3m_{\rm e} c^2\Gamma_{\rm j})/(4\sigma_{\rm T} R_{\rm j} U^\prime_{B})$ is the Lorentz factor of hypothetical electrons cooling via synchrotron radiation only.\\
We introduce two micro-physical parameters: the fraction of shock downstream energy density carried by the magnetic field, $\epsilon_{B} =B^{\prime 2} R_{\rm j}^{2} \Gamma_{\rm j}^{2} c/(2L_{\rm j})$, and the fraction carried by relativistic electrons, $\epsilon_e \simeq 0.1$, with a total power in non-thermal electrons given by $L_{\rm e} = \epsilon_{\rm e} L_{\rm j}$.\\
As a first attempt, we assume the dissipation site to be above the kilonova photosphere, whose radius is $R_\mathrm{KN} \sim 6 \times 10^{13}  \, v_\mathrm{max,ej,-0.7}\ T_{4}\, \rm cm$. In this configuration, where $R_{\rm j} > R_\mathrm{KN}$ (scenario \textbf{A}), most of the seed photons from the kilonova are de-beamed as seen in the jet comoving frame.\\
In that frame, the external photon energy density from the KN is given by
\begin{equation}
    U_{\rm ext}^{\prime} = \frac{L_{\rm seed}}{4 \pi R_{\rm j}^{2} c \Gamma_{\rm j}^{2}},
\end{equation}\\
where the kilonova luminosity is $L_\mathrm{\rm seed} \approx 3\times 10^{40}$ erg/s at $t=10^4$ s.\\
Assuming EIC to be the dominant cooling process, the Lorentz factor of electrons that cool via the EIC on a dynamical time-scale of $R_{\mathrm j}/(c\Gamma_{\mathrm j})$ is
\begin{equation}
\gamma_{\rm c,EIC} = \frac{3 \pi m_{\rm e} c^{3}  R_{\mathrm j} \Gamma_{\mathrm{j}}^{3}} {\sigma_{\rm T} L_{\rm seed}} \approx 3\times 10^{4} \ R_{\mathrm j,14} \ \Gamma_{\mathrm j,0.7}^{3}.
\label{eq:coolingEIC}
\end{equation}
In this scenario, it is reasonable to assume a slow-cooling regime for the injected electrons, i.e.\ $\gamma_{\rm m}<\gamma_{\rm c,EIC}$. By requiring the peak of the EIC to be at $\sim 100\ {\rm MeV}$, we get an estimate for the jet bulk Lorentz factor, namely
\begin{equation}
\Gamma_{\mathrm j} \approx 3\, L_\mathrm{seed,40.5}^{1/3} R_{\rm j,13.8}^{-1/3} \left(\frac{\nu_\mathrm{EIC}/\nu_\mathrm{seed}}{10^8}\right)^{1/6}.
\end{equation}
The observed luminosity is dominated by the electrons that cool at $R_{\mathrm j}/(c\Gamma_{\mathrm j})$, i.e. 
\begin{equation}
 L_{\rm EIC} \sim \tau_\mathrm{T} \left(\frac{\gamma_{\rm c}}{\gamma_{\rm m}}\right)^{3-p} \gamma_{\rm m}^{2} L_{\rm seed},
\end{equation}
where $\tau_\mathrm{T}$ is the Thomson optical depth at the dissipation site, which can be estimated as
\begin{equation}
\tau_\mathrm{T} \sim  \frac{\sigma_{\mathrm T} L_{\mathrm j} \chi_{\mathrm e}}{4 \pi R m_{\rm p} c^{3} \Gamma_{\mathrm j}^{3}}. 
\end{equation}\\
This returns $L_{\rm EIC} \sim 10^{45} \ L_\mathrm{seed,40.5}^{p-2} L_{\rm j,47} \ \chi_{\rm e,-1} \gamma_{\rm m,3}^{p-1} \ R_{\rm j,13.8}^{2-p} \ \Gamma_{\rm j,0.5}^{6-3p} \ \rm erg/s $, where the numerical values hereon are given for $p=2.5$.\\
If we assume the electrons at the dissipation site to be accelerated by internal shocks with a Lorentz factor contrast $\sim \Gamma$, then the fraction of internal energy carried by the accelerated electrons in the shock downstream is $\epsilon_{e} \approx 0.08 \ \gamma_{\rm m,3} \ \chi_{\rm e,-1} \ / (\Gamma_{\rm j,0.5}-1)$.\\
This configuration is able to match the excess observed in the high energy band of this source, but it requires unrealistic microphysical conditions: the requirement that the EIC dominates over synchrotron, i.e. $U^\prime_\mathrm{ext}>U^\prime_{B}$, implies
\begin{equation}
    \epsilon_{B}<\frac{L_{\rm seed}}{L_{\rm j}} < \frac{L_{\rm seed} \eta_{\rm rad}}{L_{\rm LAT}} \approx 3 \times 10^{-6},
\end{equation}
where $\eta_\mathrm{\rm rad} = \frac{L_\mathrm{\rm LAT}}{L_\mathrm{EIC}+L_\mathrm{j}} \approx \frac{L_\mathrm{LAT}}{L_\mathrm{j}} \approx 0.1$ and $U^\prime_\mathrm{B} = \epsilon_B L_j/(4\pi R^2_j c \Gamma^2_j)$. So, given the very low luminosity of the KN seed photons and the observed LAT luminosity of this GeV source, the required magnetisation of the low-power jet is very low.\\
By requiring that EIC is also dominant with respect to SSC, i.e. $U_{\rm ext}^{\prime}>U_{\rm syn}^{\prime}$, we find an even stronger constraint: the energy density in the photons produced by the synchrotron radiation is 
\begin{equation}
    U_{\rm syn}^{\prime} = \frac{L_{\rm syn}^{\prime}}{4\pi R_{\rm j}^{2} c}.
\end{equation}
By considering the cooling of electrons at $\gamma_{c}$ and taking into account the shape of the electron distribution, we obtain
\begin{equation}
    \epsilon_{B} \lesssim \frac{12 \pi m_{\rm p} c^{3} R_{\rm j} \Gamma_{\rm j}}{L_{\rm j}^{2} \chi_{\rm e} \sigma_{\rm T} (\frac{\gamma_{\rm c}}{\gamma_{\rm m}})^{3-p} \gamma_{\rm m}^2} L_{\rm seed} \approx 3 \times 10^{-10} R_{14} \Gamma_{\rm j,0.7}^{3} L_{\rm j,47}^{-2} \chi_{\rm e,-1}^{-1} \gamma_{\rm c/m,1}^{-0.5} \gamma_{\rm m,3}^{-2},
\end{equation}\\
Where $\gamma_{\rm c/m} = \gamma_{\rm c} / \gamma_{\rm m}$ and $L^\prime_{\rm syn} \simeq \tau_{\rm T} (\frac{\gamma_{\rm c}}{\gamma_{\rm m}})^{3-p} \gamma_{\rm m}^2 \epsilon_B L_\mathrm{\rm j}$.\\
One can not increase $\epsilon_B$ by increasing $R_{14} \Gamma_{j,0.7}^{3}$ because it would shift the observed peak of the EIC to higher energies (see Eq.\ \ref{eq:coolingEIC}). We notice that in this scenario one cannot invoke EIC fast cooling  (to increase $\epsilon_B$ by 2 orders of magnitude by extracting all the energy in electrons), since it would require that $\Gamma^{\prime}>\Gamma_{\rm j}$.\\
We consider a second scenario where the dissipation site is below the KN photosphere, i.e. $R_{\rm j}<R_{\rm KN}$, allowing the photons to be beamed in the jet comoving frame. In this case (scenario \textbf{B}), the external photon energy density is given by
\begin{equation}
    U_{\rm ext}^{\prime} = \frac{L_{\rm seed} \Gamma_{\rm j}^{2}}{4 \pi R_{\rm j}^{2} c}.
\end{equation}
We note that in this scenario the electrons must be in the fast cooling regime, since $\gamma_{c,EIC} \approx 10 R_{\rm j,13} \Gamma_{\rm j,1}^{-1} \ll \gamma_{\rm m}$, so that the peak of the EIC spectral energy will be provided by electrons at $\gamma_{\rm m}$, that is
\begin{equation}
\frac{\nu_{\rm EIC}}{\nu_{\rm seed}} \approx 2 \gamma_{\rm m}^{2} \Gamma_{\rm j}^{2}.
\end{equation}
By requiring that EIC dominates over synchrotron emission ($U_{\rm ext}^{\prime} >U_{B}^{\prime}$), we obtain
\begin{equation}
    \epsilon_B < \frac{L_{\rm seed}}{L_{\rm j}} \Gamma_{\rm j}^{4} \approx 3 \times 10^{-2} \Gamma_{\rm j,1}^{4} L_{\rm j,46}^{-1},
\end{equation}
where we used a lower reference jet power, because we have an energy budget of $\epsilon_{\rm e} L_{j}$ in the fast cooling regime. We adopt $\Gamma_{\rm j} \sim 10$ to provide 100 MeV peak of EIC, since $\gamma_{\rm m} \sim 1000$ for $\Gamma^{\prime} \sim$ few.\\
By requiring that EIC dominates over SSC ($U_{\rm ext}^{\prime}>U_{\rm syn}^{\prime}$), we obtain
\begin{equation}
    \epsilon_{B}<8 \times 10^{-6} R_{\rm j,13} \Gamma_{\rm j,1}^{5} L_{\rm j,46}^{-2} \chi_{\rm e,-1}^{-1} \gamma_{\rm m}^{-2}. 
\end{equation}
Therefore, also in this scenario the magnetic field in the shock downstream must be low, with $\epsilon_B \lesssim 10^{-6}$.

\bmhead{Inwards diffusion of kilonova photons towards the low-power jet}
For kilonova photons to be available for up-scattering within the low-power jet, and for them to be seen as beamed in the dissipation site comoving frame, the transverse photon diffusion time must be smaller than the dynamical time in a sizable fraction of the kilonova ejecta above $R_\mathrm{j}$. To show that this is the case, let $\tau_\perp \sim \kappa \rho(v,t) vt \theta$ be the transverse optical depth in a portion of angular size $\theta$ of the kilonova ejecta Lagrangian shell with velocity $v$. The corresponding diffusion time is $t_\mathrm{diff,\perp}=vt\theta \tau_\perp/c$. Setting this equal to the dynamical time $t$, we then obtain the angular distance $\theta_\mathrm{diff}(v,t)=\sqrt{c/(\kappa \rho(v,t)v^2 t)}$ over which transverse diffusion is effective at a time $t$ or, as a more informative product, the fraction $f_\mathrm{diff}(v,t)\sim 1-\cos(\theta_\mathrm{diff}(v,t))$ of the kilonova ejecta solid angle (assumed isotropic for simplicity) over which the transverse diffusion is effective. In our kilonova ejecta model, which follows \citesupp{grossman2014}, the density profile is given by $\rho(v,t)=\rho_0(t/t_0)^3(1-(v/v_\mathrm{max,ej})^2)^3$, where $t_0$ is a reference time and $\rho_0\sim (315/16)m_\mathrm{ej}/(4\pi v_\mathrm{max,ej}^3t_0^3)$ is the density normalization. Taking the ejecta heating rate due to the decay of unstable heavy isotopes produced by r-process nucleosynthesis to be \citesupp{Korobkin2012} $\dot\epsilon(t)\sim\dot\epsilon_0(t/10^4\,\mathrm{s})^{-1.3}$ with $\dot\epsilon_0=1.8\times 10^{11}\,\mathrm{erg\,s^{-1}\,g^{-1}}$, the luminosity that can diffuse inwards above a radius $R_\mathrm{j}$ is then obtained approximately from
\begin{equation}
    L_\mathrm{KN,in}(>R_\mathrm{j},t)\sim \dot\epsilon(t) \int_{R_\mathrm{j}/t}^{v_\mathrm{ph}(t)}4\pi t^3 v^2\rho(v,t)f_\mathrm{diff}(v,t)\mathrm{d}v,
\end{equation}
where $v_\mathrm{ph}(t)$ is the Lagrangian position of the radial photosphere, and we are following \citesupp{grossman2014} in removing the emission from above the photosphere, since the thermalization efficiency of nuclear decay products there is expected to be very low. The value of the above integral at $t=10^{4}\,\mathrm{s}$, as a function of $R_\mathrm{j}$ and assuming our best-fit parameters $m_\mathrm{ej}=0.02\,\mathrm{M_\odot}$ and $\kappa=0.6\,\mathrm{cm^2\,g^{-1}}$, is shown in Extended Data Fig.~\ref{fig:Lin} (panel a).\\
The evolution of $L_\mathrm{KN,in}$ as a function of time, assuming $R_\mathrm{j}=10^{13}\,\mathrm{cm}$, is shown in Extended Data Fig.~\ref{fig:Lin} (panel b). For comparison, we also show the total kilonova luminosity from the model. This demonstrates that most of the kilonova luminosity is available for up-scattering close to the peak, and also explains the expected time evolution of the external inverse Compton component.\\
As a final note, we neglected the presence of an expanding cocoon in between the jet and the kilonova ejecta. We note here that (1) the cocoon is composed of kilonova material, and hence will undergo the same heating due to r-process product decay as the kilonova ejecta, and (2) the cocoon is expected to be less massive and faster than the kilonova ejecta themselves, and therefore its contribution to the transverse optical depth is unimportant. Inwards diffusion of its cooling emission could contribute to the photon field available for up-scattering, but given the absence of an observed cooling emission at these times, this component is likely subdominant anyway. We therefore conservatively keep only the kilonova ejecta as our source of seed photons.

\bmhead{Time-dependent semi-analytical model for the EIC component}

The analytical investigations in the previous sections indicate that the EIC scattering of kilonova photons in a low-power jet could be a viable scenario to explain the observed excess on top of the forward shock synchrotron emission at 0.1-1 GeV at around $10^4$ s. Therefore, we set out to explore such a scenario more in detail by setting up a simple semi-analytical model that enables us to predict the light curves and spectra of such a component. We assumed the jet to remain active after the end of the prompt emission, with a decaying isotropic-equivalent kinetic luminosity $L_\mathrm{j}=L_\mathrm{j,0}(t/T)^{-a}$ and bulk Lorentz factor $\Gamma_\mathrm{j}=\Gamma_\mathrm{j,0}(t/T)^{-b}$, with $L_\mathrm{j,0}\sim 6\times 10^{51}\,\mathrm{erg/s}$ (where $T\sim 34\,\mathrm{s}$ is the prompt emission duration as measured by \textit{Fermi}/GBM and $L_\mathrm{j,0}\sim E/T$) and $\Gamma_\mathrm{j,0}\sim 1200$ (from our afterglow modelling). Based on the expectation that marginally bound material from the tidally disrupted least-massive neutron star in the progenitor binary will keep falling back \citesupp{Rosswog2007} and feed an accretion disk with an accretion rate $\dot M\propto t^{-5/3}$ \citesupp{Rees1988}, we set $a=5/3$, while $b>0$ remains as a free parameter. Under these assumptions, we can compute at any time the radius of the jet photopshere from \citesupp{daigne2002} $R_\mathrm{ph}\sim \kappa_\mathrm{es}L_\mathrm{j}/(8\pi \Gamma_\mathrm{j}^3c^3)\sim 10^9\,(t/T)^{3b-5/3}\,\mathrm{cm}$ where we took the electron scattering opacity $\kappa_\mathrm{es}=0.2\,\mathrm{cm^2\,g^{-1}}$, appropriate for a $Y_\mathrm{e}=1/2$ electron fraction. Given the very large Lorentz factor and relatively small luminosity, the photosphere therefore starts off very deep in the jet. We assume dissipation events (specifically, internal shocks) to happen randomly throughout the jet from the photosphere up, with an optimistic Lorentz factor contrast $\Gamma'\sim\Gamma_\mathrm{j}$, so that electron are accelerated with a power law energy distribution $\mathrm{d}n_\mathrm{e}/\mathrm{d}\gamma\propto \gamma^{-p}$ at the dissipation site above an injection Lorentz factor $\gamma_\mathrm{m}=1 + (p-2)/[(p-1)(\epsilon_\mathrm{e}/\chi_\mathrm{e})(m_\mathrm{p}/m_\mathrm{e})(\Gamma_\mathrm{j}-1)]$, within an isotropic-equivalent comoving volume $V'\sim 4\pi R_\mathrm{j}^3/\Gamma_\mathrm{j}$. 
We assume the dominant EIC emission to come from kilonova photons upscattered at a radius $R_\mathrm{j}\sim\max(R_\mathrm{ph},R_\mathrm{KN})$, where the latter quantity indicates the kilonova photosphere. Whenever $R_\mathrm{ph}<R_\mathrm{KN}$, we assume the kilonova photons to have energies doppler-boosted by a factor $\Gamma_\mathrm{j}$ in the comoving frame of the jet, so that their energy density in that frame, at the dissipation site, is $U'_\mathrm{ext}=\Gamma_\mathrm{j}^2L_\mathrm{KN}/(4\pi R_\mathrm{j}^2c)$. Conversely, if $R_\mathrm{ph}>R_\mathrm{KN}$, then we assume the photons to be de-beamed, so that $U'_\mathrm{ext}=\Gamma_\mathrm{j}^{-2}L_\mathrm{KN}/(4\pi R_\mathrm{j}^2c)$. Accordingly, the kilonova temperature, as seen in the jet comoving frame is $T'_\mathrm{ext}=\Gamma_\mathrm{j}^s T_\mathrm{KN}$, with $s=1$ if $R_\mathrm{ph}<R_\mathrm{KN}$ and $s=-1$ otherwise. 
The relativistic electrons in the jet dissipation site emit (and cool) by synchrotron, SSC and EIC: the cooling Lorentz factor $\gamma_\mathrm{c}$, at which the cooling time equals the dynamical time, is obtained by solving numerically the implicit equation
\begin{equation}
    \gamma_\mathrm{c}(1+Y_\mathrm{EIC}+Y_\mathrm{SSC}(\gamma_\mathrm{c}))=\frac{6\pi m_\mathrm{e}c^2 \Gamma_\mathrm{j}}{\sigma_\mathrm{T}B^{\prime 2} R_\mathrm{j}},
\end{equation}
where $Y_\mathrm{EIC}=U'_\mathrm{ext}/U'_{B}$ and $Y_\mathrm{SSC}(\gamma_\mathrm{c})\sim n'_\mathrm{e} P_\mathrm{\nu,syn,max}\nu'_0/U'_{B}$, with $P_\mathrm{\nu,syn,max}=\sigma_\mathrm{T}m_\mathrm{e}c^2B^\prime/(3e)$, $n'_\mathrm{e}\sim \chi_\mathrm{e}L_\mathrm{j,iso}/(4\pi R_\mathrm{j}^2\Gamma_\mathrm{j}^2 m_\mathrm{p} c^3)$ and $\nu'_0 = \max(\gamma_\mathrm{m}^2,\gamma_\mathrm{c}^2)eB^\prime/(2\pi m_\mathrm{e}c)$, and assuming the usual parametrization $U'_{B} = \epsilon_{B}L_\mathrm{j,iso}/(4\pi R_\mathrm{j}^2\Gamma_\mathrm{j}^2c)$. All these assumptions completely specify the electron and photon energy distributions in the comoving frame. Given these, we compute the EIC emissivity in the low-power jet as $j'_\mathrm{\nu,EIC}=[h\nu'/(4\pi)][\mathrm{d^2}n^\prime_\gamma/(\mathrm{d}\nu\,\mathrm{d}t)]$, where the comoving density of photons emitted per unit time is
\begin{equation}
\frac{\mathrm{d^2}n^\prime_\gamma}{\mathrm{d}\nu\,\mathrm{d}t} = \int_{\min(\gamma_\mathrm{m},\gamma_\mathrm{c})}^\infty \frac{\mathrm{d^2}N_\mathrm{\gamma,EIC}}{\mathrm{d}\nu\mathrm{d}t}(\nu',T_\mathrm{ext}^\prime,\kappa_\mathrm{d})\frac{\mathrm{d}n_e^\prime}{\mathrm{d}\gamma}\mathrm{d}\gamma.
\end{equation}
We compute the external inverse Compton specific photon scattering rate per electron $\mathrm{d^2}N_\mathrm{\gamma,EIC}/(\mathrm{d}\nu\,\mathrm{d}t)$ adopting the simple analytical approximation from \citesupp{Khangulyan2014} (their Eq.~14), using a dilution factor $\kappa_\mathrm{d}=\Delta\Omega/(4\pi)\sim[R_\mathrm{j}/(2R_\mathrm{KN})]^2=1/4$ (this is a conservative choice, as the actual kilonova solid angle $\Delta\Omega$, as seen in the jet comoving frame, is possibly larger than $\pi$) and a temperature $T'_\mathrm{ext}=\Gamma_\mathrm{j}^s T_\mathrm{KN}$. The electron Lorentz factor distribution is assumed to have the usual form \citesupp{Panaitescu2000}
\begin{equation}
    \frac{\mathrm{d}n_\mathrm{e}^\prime}{\mathrm{d}\gamma}\propto\left\lbrace\begin{array}{lr}
    (\gamma/\gamma_0)^{-q}  & \gamma_\mathrm{p}\leq \gamma \leq \gamma_0  \\
    (\gamma/\gamma_0)^{-p-1} & \gamma > \gamma_\mathrm{0}
    \end{array}\right.,
\end{equation}
where $\gamma_\mathrm{p}=\min(\gamma_\mathrm{m},\gamma_\mathrm{c})$, $\gamma_0=\max(\gamma_\mathrm{m},\gamma_\mathrm{c})$, $q=p$ if $\gamma_\mathrm{m}\leq\gamma_\mathrm{c}$ and $q=2$ otherwise. The normalization is such that $\int (\mathrm{d}n_\mathrm{e}^\prime/\mathrm{d}\gamma)\mathrm{d}\gamma = N_\mathrm{e}/V^\prime = \chi_\mathrm{e}L_\mathrm{j}/(4\pi R_\mathrm{j}^2\Gamma_\mathrm{j}^2 m_\mathrm{p}c^3)$.
We then compute the final EIC specific luminosity as 
\begin{equation}
\frac{\mathrm{d}L_\mathrm{EIC}}{\mathrm{d}\nu}=\Gamma_\mathrm{j}^2 V' j'_\mathrm{\nu,EIC}f_\mathrm{jet},    
\end{equation}
where $f_\mathrm{jet}=\min(1,\Gamma_\mathrm{j}^2\theta_\mathrm{j}^2)$ accounts for the reduction in the observed luminosity when $\Gamma_\mathrm{j}<\theta_\mathrm{j}^{-1}$, an we use $\theta_\mathrm{j}=1\,\mathrm{deg}$ as found from the afterglow modelling.\\
The resulting model has five free parameters, namely $b$, $\epsilon_\mathrm{e}$, $\epsilon_{B}$, $p$ and $\chi_\mathrm{e}$. We found that $b=0.6$, $\epsilon_\mathrm{e}=\chi_\mathrm{e}=0.1$, $\epsilon_\mathrm{B}=10^{-8}$ and $p=2.2$ provides a viable solution, that reproduces both the LAT light curve and the SEDs (see Fig.~\ref{fig:grb1_sed_lc_EIC} in the main text), with the jet synchrotron emission remaining below the forward shock one, as shown in Extended Data Fig.~\ref{fig:EIC+jetSYN}.\\
The evolution of the EIC component shown in Fig.~\ref{fig:grb1_sed_lc_EIC} is the combined result of the evolution of the jet power and Lorentz factor, the kilonova luminosity and temperature, and the energy density of the kilonova photons as seen in the jet comoving frame at the dissipation radius. The latter determines the contribution of EIC to the cooling of electrons accelerated at the dissipation region, which affects the shape of the effective electron energy distribution that produces the EIC emission upon scattering of the kilonova photons.

\bmhead{Comparison with afterglow models from other works}
GRB 211211A has been analysed by several other groups \citesupp{Rastinejad2022,Xiao2022,yang2022}. The preferred afterglow model parameters reported in these works are different from ours. To address this discrepancy, we produced the same diagnostic plots as in Fig.~\ref{fig:grb1_sed_lc}, but using their best-fit parameters, and adopting the \texttt{afterglowpy} \citesupp{Ryan2020} software (which allows for off-axis viewing angles and structured jets, as opposed to our modelling which assumes the line of sight to be on the jet axis, and the jet properties to be uniform within the jet opening angle). As shown in Extended Data Figures~\ref{fig:rastinejad}, \ref{fig:yang} and \ref{fig:xiao}, the parameters reported in the preceding works typically lead to predictions that match the multi-wavelength light curves at $t\gtrsim 10^4\,\mathrm{s}$, but fail to reproduce the early optical data and the XRT photon index (except for \citesupp{Xiao2022}, whose model has the correct photon index in the XRT band). Similarly to our model, the model from \citesupp{Xiao2022} produces a similar flux as the observed one in the LAT band in correspondence of the first detection, but with an inconsistent photon index. All models (including ours, see Fig.~\ref{fig:grb1_sed_lc}) instead are well below the second LAT detection, further supporting its interpretation as an excess over the synchrotron afterglow.\\
Recently, \citesupp{zhang2022} independently analyzed the \lat\ data and confirmed the GeV detection of GRB 211211A. Their flux values are similar to the ones estimated in the present work. However, there are significant  differences on data and modelling as detailed in the following. Their Figure 3 shows the first upper limit starting at 400 s and covering about 1.5 ks. 
However, up to 1 ks the source was off-axis ($\theta >$ 65 deg) with respect to the LAT FoV and no upper limit can be reliably estimated in this interval. 
In addition to that, 
their interpretation, which attributes the late high-energy emission to synchrotron radiation from the forward shock,
uses only an arbitrary subset of the available data, especially in the UVOIR bands. This, together with the fact that their numerical afterglow model is not publicly available (and we are not able to reproduce their result using our own model, which could be attributed to differences in the details of the implementation) makes it difficult to assess independently whether their interpretation is really viable (we note, however, that their initial Lorentz factor is in a region with very little posterior support according to our results, see Extended Data Figure \ref{fig:afterglow_model_corner_plot}). In that case, the early Optical observation by UVOT would then have to be explained separately.

\bibliographystylesupp{naturemag}
\bibliographysupp{references}

\section*{Declarations}

\bmhead{Acknowledgments}
We thank Annalisa Celotti, Gabriele Ghisellini, Alberto Segreto, Elena Ambrosi and Maria Grazia Bernardini for fruitful discussions. The National Radio Astronomy Observatory is a facility of the National Science Foundation operated under cooperative agreement by Associated Universities, Inc. PDA and SC warmly thank Norbel Schartel for granting {\it XMM}-Newton DDT observations.
MB acknowledge financial support from the Italian Ministry of University and Research (MUR, PRIN 2020 grant 2020KB33TP). MB and GO acknowledge financial support from the AHEAD2020 project (grant agreement n. 871158). BB, MB and PDA acknowledge financial support from MUR (PRIN 2017 grant 20179ZF5KS). OS thanks MUR grant 2017MB8AEZ for financial support. PDA and SC acknowledge support from ASI grant I/004/11/5. This work made use of data supplied by the UK Swift Science Data Centre at the University of Leicester.

\bmhead{Author contribution statement}
A.M. , B.B. and G.O. carried out LAT data reduction and analysis, and lead the GeV discovery. O.S. performed the multi-wavelength afterglow modelling. G.O. and O.S. developed the theoretical model used to interpret the high energy excess. A.M. , B.B., G.O., O.S. lead the paper writing. G.G., M.B., S.G., P.D.A. and S.R. gave significant inputs on data interpretation. M.B. and G.G. gave major contributions to the paper writing. All authors contributed to discussions and editing of the paper. P.D.A. is the principal investigator of the XMM-Newton observations, and collected and analysed the optical and UV data. P.D.A. and S.C. reduced and analysed the XMM data and edited the corresponding text in the paper. S.G. is the principal investigator of the VLA observations. He reduced and analysed the radio data, and edited the corresponding text in the paper. P.T. and A.S. contributed to the LAT analysis providing computational tools. B.B. produced Fig. 1 and Extended Data Fig. 1. A.M. produced Fig. 2 and Extended Data Fig. 2. O.S. produced Fig. 3, Fig.4  and Extended Data Fig. 3, 4, 5, 6, 7, 8. S.R. produced Fig. 5.

\bmhead{Author information statement}
Correspondence and requests for materials should be addressed to Alessio Mei. 

\bmhead{Conflict of interest}
We declare no conflicts of interest.

\bmhead{Data availability}
\xrt\ raw data are public and available in the UK Swift Science Data Centre at the University
of Leicester. The light curve data are taken at this link:\\
\texttt{https://www.swift.ac.uk/xrt\textunderscore curves/GRB\textunderscore ID/flux.qdp}\\
where \texttt{GRB\textunderscore ID} is the GRB observation ID.\\
The spectra are obtained at this link:\\
\texttt{https://www.swift.ac.uk/xrt\textunderscore spectra/addspec.php?targ=GRB\textunderscore ID}\\
where \texttt{GRB\textunderscore ID} is the GRB observation ID.\\
The details of the automatic spectral analysis can be found here:\\
\texttt{https://www.swift.ac.uk/xrt\textunderscore spectra/docs.php}\\
\textit{Swift}/UVOT raw data are available at the link:\\
\texttt{https://heasarc.gsfc.nasa.gov/cgi-bin/W3Browse/swift.pl}\\
\lat\ raw data are public and can be downloaded using the software \textsc{gtburst}:\\
\texttt{https://fermi.gsfc.nasa.gov/ssc/data/analysis/scitools/gtburst.html}\\
\lat\ 2nd GRB catalog data are available at the following link:
\texttt{https://www-glast.stanford.edu/pub\textunderscore data/953/}\\
VLA data are available at the public repository:\\
{\tt https://data.nrao.edu/portal/\#/}\\
The observation code is 21B-370.\\
XMM-Newton raw data are available at the link:\\
{\tt https://www.cosmos.esa.int/web/xmm-newton/xsa}\\
TNG data are available from the corresponding author upon reasonable request.

\bmhead{Code availability}

\textsc{Heasoft}, \textsc{XSPEC} and \textsc{PyXspec} are freely available online at the following links:\\
\texttt{https://heasarc.gsfc.nasa.gov/docs/software/heasoft/}\\
\texttt{https://heasarc.gsfc.nasa.gov/xanadu/xspec/}\\
\texttt{https://heasarc.gsfc.nasa.gov/docs/xanadu/xspec/python/html/index.html}\\
\textsc{gtburst} is one of the Fermi Science Tools package, freely available at the following link:\\
\texttt{https://fermi.gsfc.nasa.gov/ssc/data/analysis/software/}\\
The details of the \textsc{gtburst} analysis can be found here:\\
\texttt{https://fermi.gsfc.nasa.gov/ssc/data/analysis/scitools/gtburst.html}\\
\textsc{SAS} is freely available at the following link:\\
\texttt{https://www.cosmos.esa.int/web/xmm-newton/sas-download}\\
\textsc{Sosta} is part of the \textsc{Ximage} software package, freely available at:\\
\texttt{https://heasarc.gsfc.nasa.gov/xanadu/ximage/ximage.html}\\
The \textsc{ESO-eclipse} package is available at the link:\\
\texttt{https://www.eso.org/sci/software/eclipse/}\\
\textsc{casa} software is available at the following link:\\
\texttt{https://casa.nrao.edu/casa\textunderscore obtaining.shtml}\\
A tutorial on how to use \textsc{casa} can be found at the following link:\\
{\tt https://casaguides.nrao.edu/index.php?title=VLA\textunderscore Continuum\textunderscore Tutorial\\\textunderscore 3C391-CASA6.2.0}\\
\textsc{emcee} is a python package, available at the following link:\\
\texttt{https://emcee.readthedocs.io/en/stable/user/install/}\\
\textsc{afterglowpy} is a python package, available at the following link:\\
\texttt{https://github.com/geoffryan/afterglowpy}\\
All reduced data and computer code are available from the corresponding authors upon reasonable request.\\
\\
Reprints and permissions information is available at:\\
{\tt www.nature.com/reprints.}

\section*{Extended Data}\label{sec:methods}

\captionsetup[table]{name = Extended Data Table }
\captionsetup[figure]{name = Extended Data Figure }

\begin{table}[h] 
\caption{}
    \centering
    \begin{tabular}{ |c | c | c | c|} \hline
Energy (GeV) & Probability & Distance (deg.) & Arrival time (sec.) \\ \hline
0.21 & 	0.94 & 	0.36 & 	6438.18 \\ 
0.19 & 	0.95 & 	1.04 & 	6647.43 \\ 
0.16 & 	0.93 & 	1.34 & 	12493.41 \\ 
0.12 & 	0.96 & 	0.71 & 	12612.52 \\ 
1.74 & 	0.97 & 	0.32 & 	12966.74 \\ 
0.10 & 	0.96 & 	0.77 & 	13053.43 \\ 
0.12 & 	0.92 & 	1.69 & 	13292.13 \\ 
0.29 & 	0.91 & 	1.22 & 	17860.45 \\ 
0.23 & 	0.97 & 	0.67 & 	18127.51 \\ \hline
    \end{tabular}
    \label{tab:GeVPhotons}
\end{table}

\begin{table}[h]
\caption{}
    \centering
    \begin{tabular}{ccc}
        Parameter & Posterior$^a$ & Prior bounds \\
        \hline
        $\log(E/\mathrm{erg})$ & $53.2^{+0.8}_{-1.0}$ & $(50,55)$\\
         $\log(n/\mathrm{cm^{-3}})$ & $<-4.2$ & $(-6,2)$ \\
         $\log(\Gamma_0)$ & $3.1^{+0.9}_{-0.6}$ & $(1,4)$ \\
         $\log(\theta_j/\mathrm{rad})$ & $-1.74^{+0.18}_{-0.19}$ & $(-2,0)$ \\
         $\log\epsilon_\mathrm{e}$ & $-1.6^{+1.0}_{-0.67}$ & $(-3,-0.3)$ \\
         $\log\epsilon_\mathrm{B}$ & $-2.5^{+1.1}_{-1.0}$ & $(-7,-0.3)$ \\
         $\log\chi_\mathrm{e}$ & $ <-0.52$ & $(-2,0)$ \\
         $p$ & $2.31^{+0.14}_{-0.10}$ & $(2.01,2.99)$\\
         $\log(m_\mathrm{ej}/\mathrm{M_\odot})$ & $-1.7^{+0.17}_{-0.17}$ & $(-4,0)$ \\
         $\log(\kappa/\mathrm{cm^2\,g^{-1}})$ & $-0.21^{+0.36}_{-0.31}$ & $(-1,2)$ \\
         $\log(v_\mathrm{max,ej}/c)$ & $-0.71^{+0.25}_{-0.24}$ &  $(-2,-0.2)$ \\
         $\log(f_\mathrm{sys})$ & $-0.77^{+0.20}_{-0.21}$ & $(-5,0)$\\
         \hline
    \end{tabular}\\
    \footnotesize{$^a$Best fit value (median of posterior samples) and 90\% credible range (or 95\% credible upper limit) constructed from marginalised posterior.}
    \label{tab:model_fitting}
\end{table}

\begin{figure*}[h]
    \centering 
    \setcounter{figure}{0}    

    \includegraphics[width=0.7\textwidth]{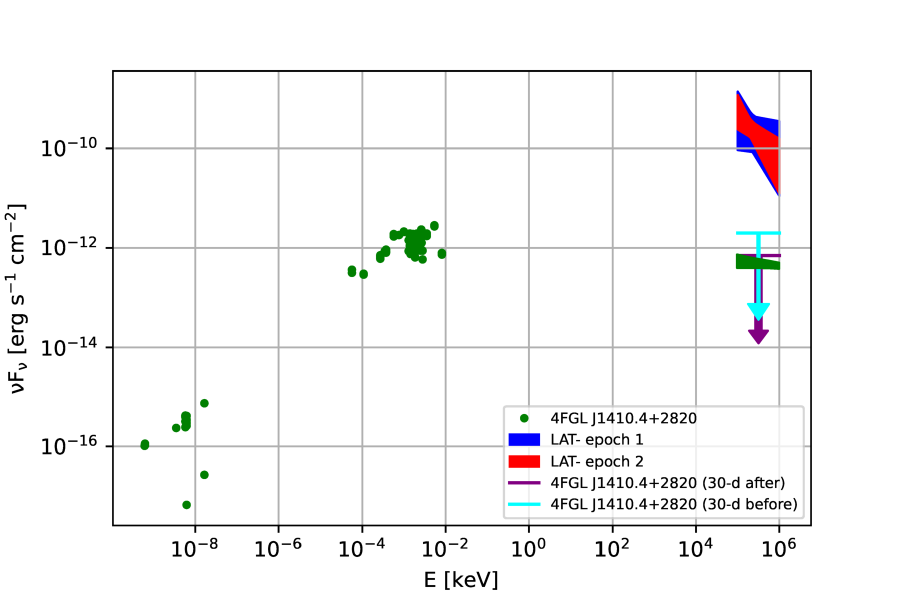} 
    \caption{}
    \label{fig:blazar_sed}
\end{figure*}

\begin{figure*}[h]
	\centering 	
	\includegraphics[width=1\textwidth]{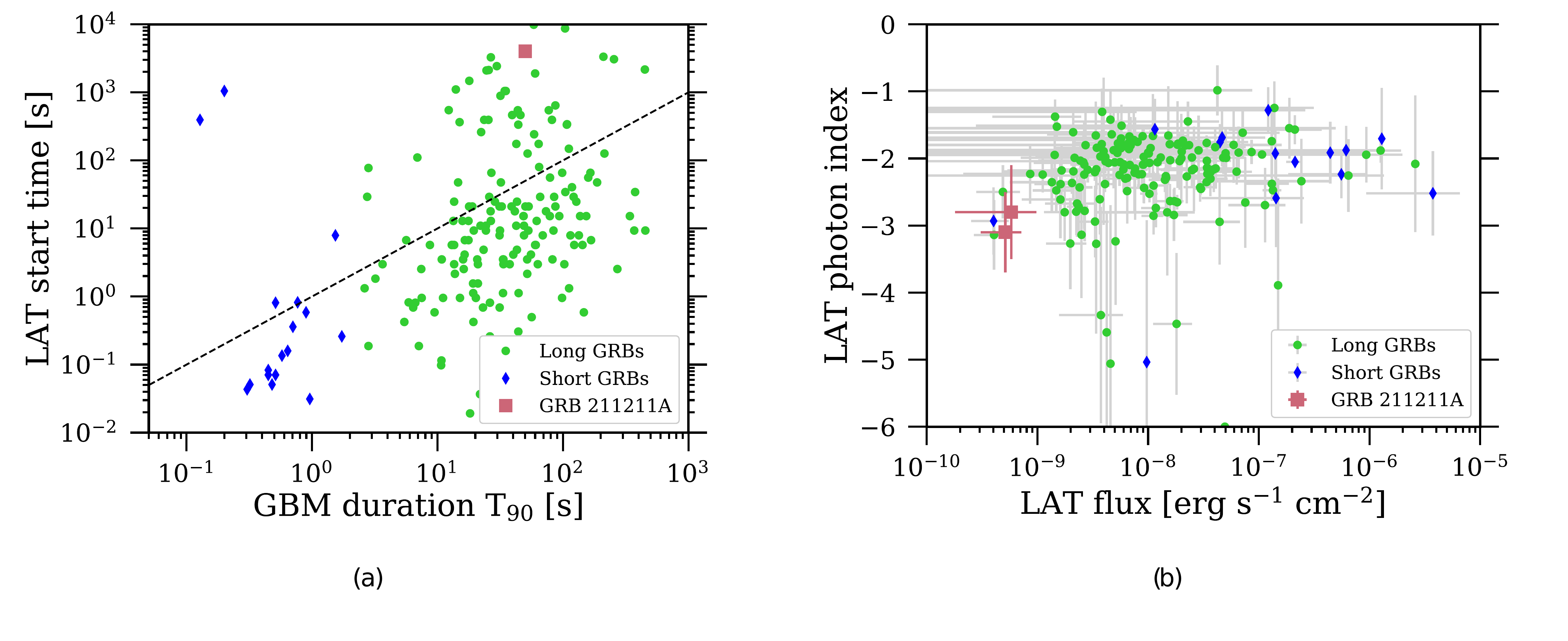} 
	\caption{}
	\label{fig:LAT_catalog2}
\end{figure*}

\begin{figure*}[h]
    \centering
    \includegraphics[width=\textwidth]{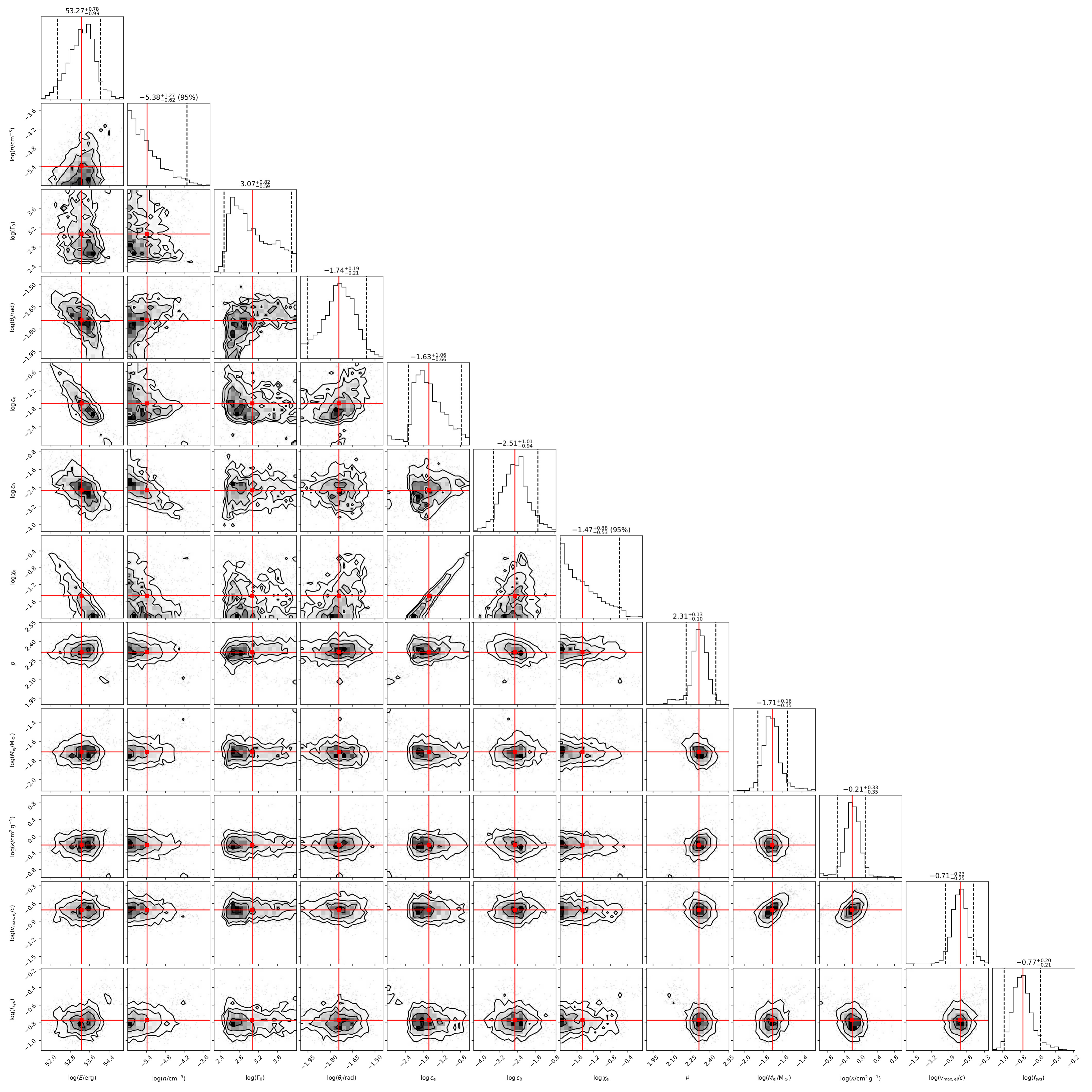}
    \caption{}
    \label{fig:afterglow_model_corner_plot}
\end{figure*}

\begin{figure*}[h]
	\centering 	
	\includegraphics[width=1\textwidth]{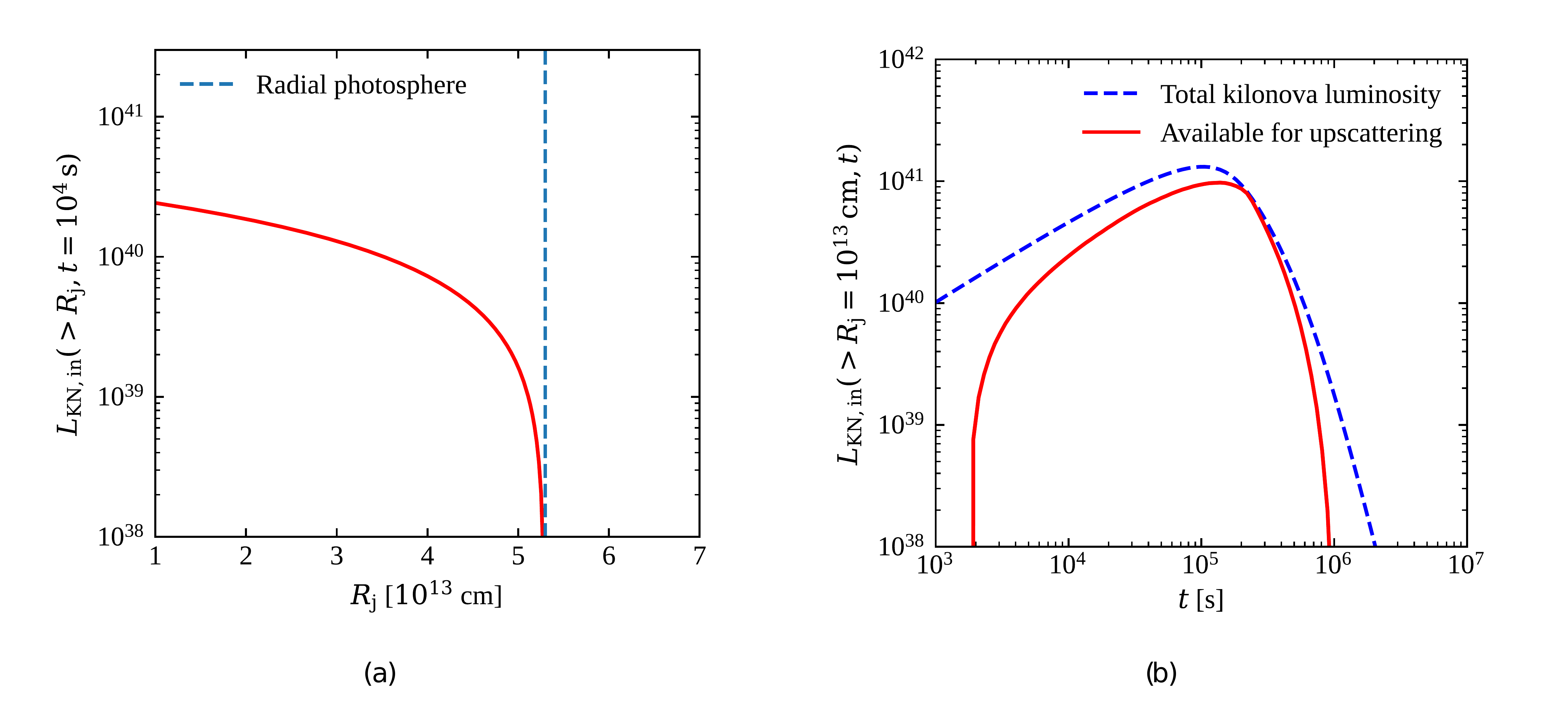} 
	\caption{}
	\label{fig:Lin}
\end{figure*}

\begin{figure}
    \centering
    \includegraphics[width=1\textwidth]{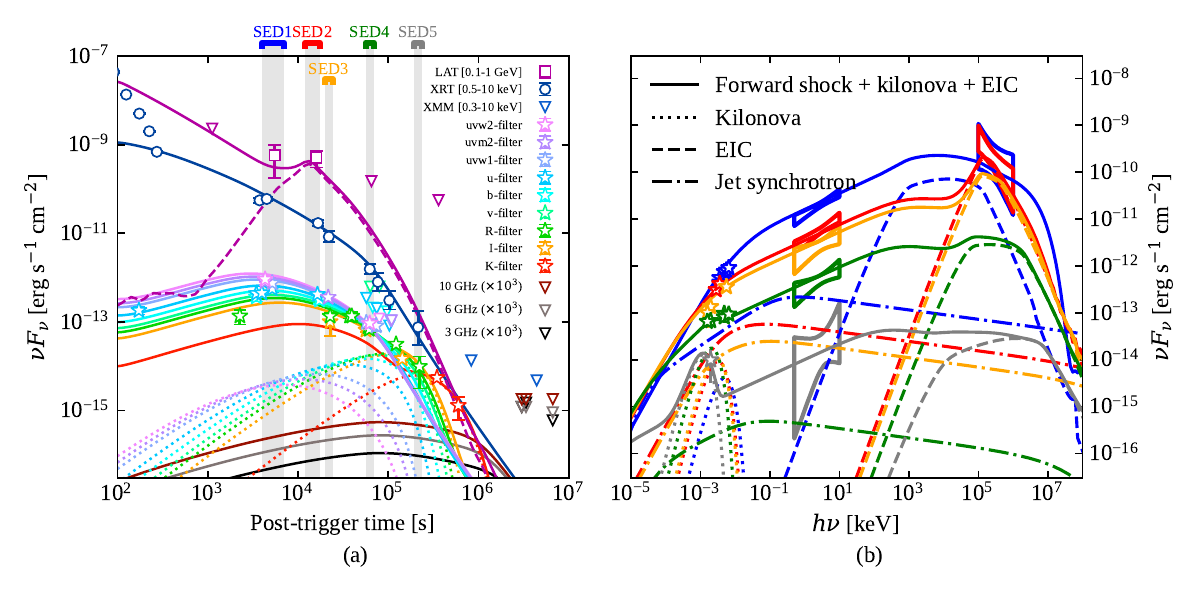}
    \caption{}
    \label{fig:EIC+jetSYN}
\end{figure}

\begin{figure*}[h]
    \centering
    \includegraphics[width=\textwidth]{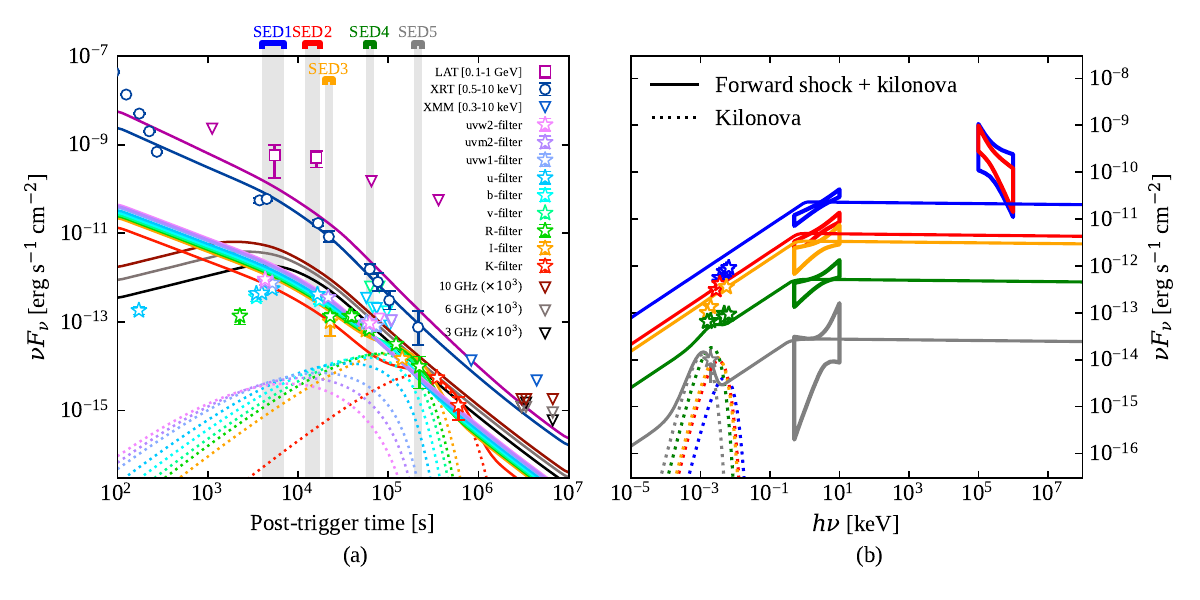}
    \caption{}
    \label{fig:rastinejad}
\end{figure*}

\begin{figure*}[h]
    \centering
    \includegraphics[width=\textwidth]{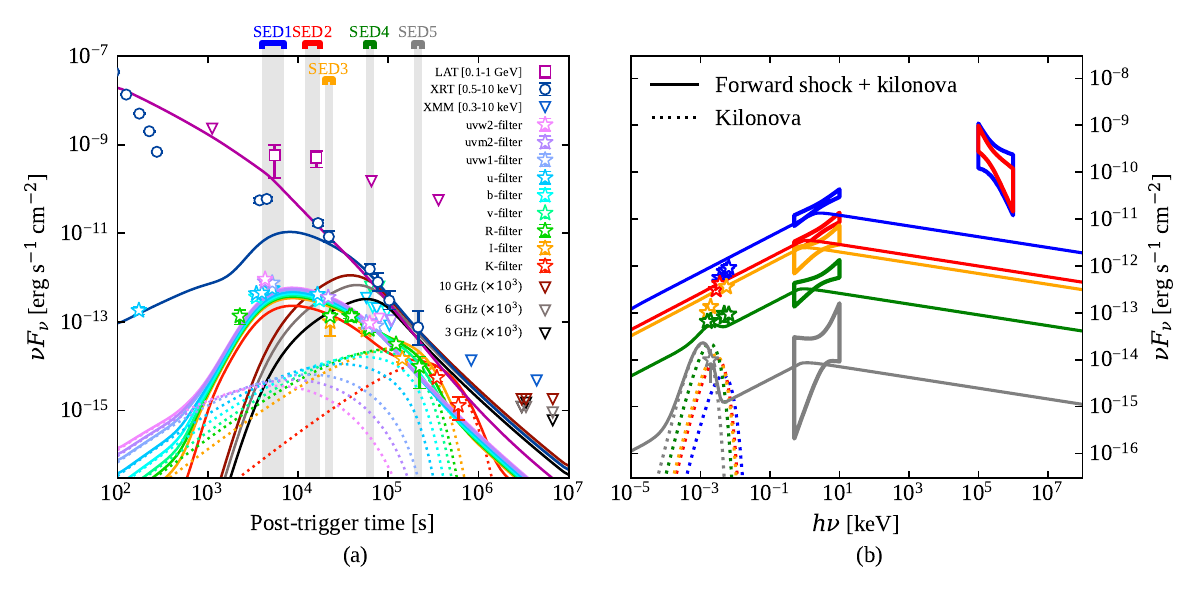}
    \caption{}
    \label{fig:yang}
\end{figure*}

\begin{figure*}[h]
    \centering
    \includegraphics[width=\textwidth]{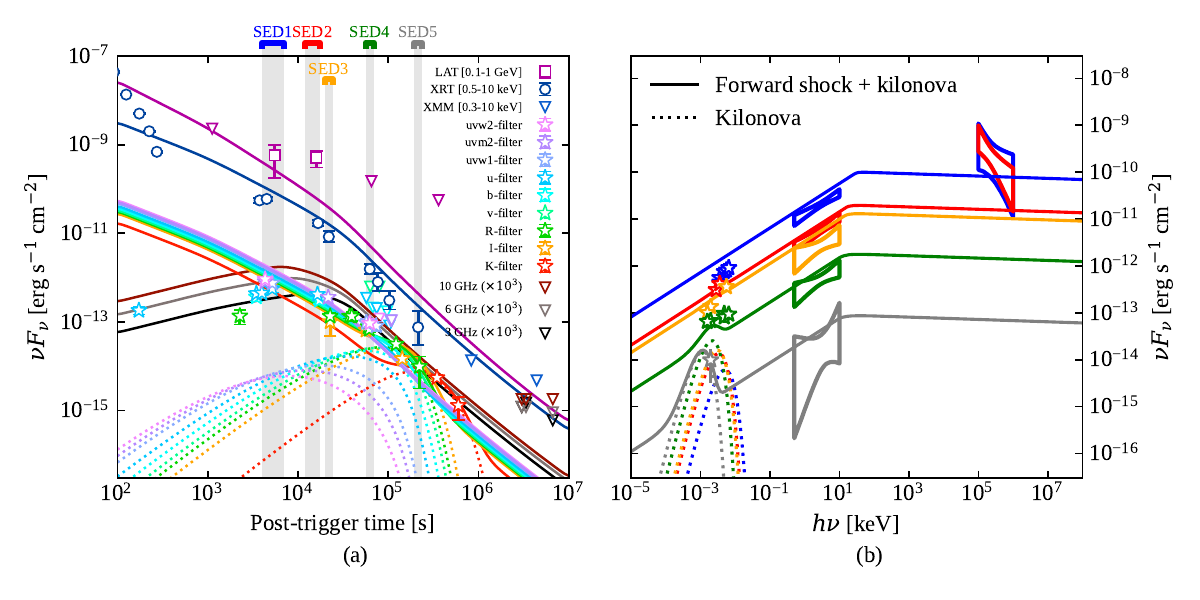}
    \caption{}
    \label{fig:xiao}
\end{figure*}

\clearpage

\bmhead{Extended Data Table 1: High energy photons from GRB 211211A detected by \lat}
The energy of the photons, probability of association to the GRB, distance from the GRB location (as respect to RA=212.272$^\circ$, DEC=27.884$^\circ$) and the arrival time of the photons from the trigger time with probability more than 0.9 .

\bmhead{Extended Data Table 2: Results of the forward shock + kilonova model fit}
Parameters of our forward shock + kilonova model, summary results from our MCMC analysis, and bounds of the adopted priors.

\bmhead{Extended Data Figure 1: Time-averaged broadband spectrum of 4FGL J1410.4+2820}
The two arrows represent the 3$\sigma$ upper limits for the BL Lac flux obtained using one month of observation by {\it Fermi}/LAT before and after the GRB (in yellow and purple, respectively). The green band in the GeV energies represents the time-averaged GeV emission from 12 years of observation \citesupp{2022arXiv220111184F}. The emission from the blazar is at least two orders of magnitude weaker than the emission from the GRB.

\bmhead{Extended Data Figure 2: Comparisons with other GRBs observed by \lat}
Long (in green) and short (in blue) bursts emissions from the 2nd \lat\ GRB catalog \cite{ajello2019} compared to GRB 211211A (in brown). We show the LAT detection time from the burst versus the GRB duration $\rm T_{90}$ (a) computed with \gbm\ data. The dashed lines separate GRBs that are detected during (below) or after (above) the prompt emission. Note that in some cases, including GRB 211211A, the {\it Fermi}/LAT observation started after the prompt phase, and we cannot exclude an emission starting before the detection time shown in the plot. In panel (b) we show LAT photon index versus LAT flux (0.1 - 10 GeV), both obtained through time-integrated analysis in \cite{ajello2019}.

\bmhead{Extended Data Figure 3: Corner plot of the 12-dimensional posterior obtained from MCMC sampling}
The meaning of the parameters is explained in the text. The histograms on the diagonal show the one-dimensional marginalised posterior probability density for each parameter, with the red line showing the best fit and the dashed lines bracketing 90\% (or 95\% in case of upper limits) credible ranges. Contours in the remaining two-dimensional plots show the one, two and three-sigma equivalent bounds of the joint posteriors of parameter pairs, while dots show qualitatively the distribution of posterior samples outside the three-sigma boundaries. The red lines and dots show the position of the best fit.

\bmhead{Extended Data Figure 4: Details on the kilonova photon transverse diffusion}
Kilonova luminosity that can diffuse from the jet-kilonova `walls' above the jet dissipation region, located at $R_\mathrm{j}$, at post-merger time $t=10^{4}$ s (a). In panel (b), we show the Kilonova luminosity available for up-scattering within the jet dissipation region (red solid line), compared to the total kilonova luminosity (blue dashed line), assuming $R_\mathrm{j}=10^{13}\,\mathrm{cm}$

\bmhead{Extended Data Figure 5: Light curves as SEDs with models, showing the synchrotron emission from the low-power jet}
Same as Figure \ref{fig:grb1_sed_lc}, but showing the model-predicted synchrotron emission from the low-power jet with dot-dashed lines in panel (b).

\bmhead{Extended Data Figure 6: Comparison with the afterglow modelling in Rastinejad et al.\ 2022}
Light curves (a) and SEDs (b) with the best fit parameters from Rastinejad et al.\ 2022 \citesupp{Rastinejad2022}.

\bmhead{Extended Data Figure 7: Comparison with the afterglow modelling in Yang et al.\ 2022}
Light curves (a) and SEDs (b) with the best fit parameters from Yang et al.\ 2022 \citesupp{yang2022}.

\bmhead{Extended Data Figure 8: Comparison with the afterglow modelling in Xiao et al.\ 2022}
Light curves (a) and SEDs (b) with the best fit parameters from Xiao et al.\ 2022 \citesupp{Xiao2022}.

\appendix

\end{document}